\documentclass[aps,twocolumn,english,superscriptaddress,citeautoscript,showkeys,preprintnumbers,amsmath,amssymb,floatfix,footinbib,prb]{revtex4-2}
\usepackage{xcolor}
\usepackage{soul}
\usepackage{amssymb,amsmath}
\usepackage{hyperref}
\hypersetup{colorlinks=true,linkcolor=red,citecolor=blue}
\usepackage[utf8]{inputenc}
\usepackage[english]{babel}
\usepackage{txfonts}

\usepackage{graphicx} 
\usepackage{dcolumn} 
\usepackage{bm} 
\usepackage{float}
\usepackage{hyperref}

\begin{document}

\title{Nodeless time-reversal symmetry breaking in the centrosymmetric superconductor Sc$_5$Co$_4$Si$_{10}$ probed by muon-spin spectroscopy}

\author{A. Bhattacharyya}
\email{amitava.bhattacharyya@rkmvu.ac.in} 
\address{Department of Physics, Ramakrishna Mission Vivekananda Educational and Research Institute, Belur Math, Howrah 711202, West Bengal, India}
\author{M. R. Lees}
\affiliation{Department of Physics, University of Warwick, Coventry CV4 7AL, United Kingdom}
\author{K. Panda}
\address{Department of Physics, Ramakrishna Mission Vivekananda Educational and Research Institute, Belur Math, Howrah 711202, West Bengal, India}
\author{P. P. Ferreira}
\email{pedroferreira@usp.br}
\affiliation{Computational Materials Science Group (ComputEEL/MatSci), Escola de Engenharia de Lorena, Universidade de São Paulo, DEMAR, Lorena, Brazil} 
\author{T. T. Dorini}
\affiliation{Université de Lorraine, CNRS, IJL, Nancy, France}
\author{Emilie Gaudry}
\affiliation{Université de Lorraine, CNRS, IJL, Nancy, France}
\author{L. T. F. Eleno}
\affiliation{Computational Materials Science Group (ComputEEL/MatSci), Escola de Engenharia de Lorena, Universidade de São Paulo, DEMAR, Lorena, Brazil}
\author{V. K. Anand}
\affiliation{Helmholtz-Zentrum Berlin f\"{u}r Materialien und Energie GmbH, Hahn-Meitner Platz 1, D-14109 Berlin, Germany
}
\affiliation{Department of Physics, University of Petroleum and Energy Studies, Dehradun, Uttarakhand 248007, India}
\author{J. Sannigrahi}
\affiliation{School of Physical Sciences, Indian Institute of Technology Goa, Goa-403401, India}
\author{P. K. Biswas} 
\affiliation{ISIS Neutron and Muon Source, Rutherford Appleton Laboratory, Chilton, Didcot Oxon, OX11 0QX, United Kingdom} 
\author{R. Tripathi}
\affiliation{ISIS Neutron and Muon Source, Rutherford Appleton Laboratory, Chilton, Didcot Oxon, OX11 0QX, United Kingdom}
\affiliation{Jawaharlal Nehru Centre for Advanced Scientific Research, Jakkur, Bangalore 560064, India}
\author{D. T. Adroja} 
\email{devashibhai.adroja@stfc.ac.uk}
\affiliation{ISIS Neutron and Muon Source, Rutherford Appleton Laboratory, Chilton, Didcot Oxon, OX11 0QX, United Kingdom} 
\affiliation{Highly Correlated Matter Research Group, Physics Department, University of Johannesburg, PO Box 524, Auckland Park 2006, South Africa}
 
\date{\today}
\begin{abstract}
We investigate the superconducting properties of Sc$_{5}$Co$_{4}$Si$_{10}$ using low-temperature resistivity, magnetization, heat capacity, and muon-spin rotation and relaxation ($\mu$SR) measurements. We find that Sc$_{5}$Co$_{4}$Si$_{10}$ {exhibits type-II} superconductivity with a superconducting transition temperature $T_\mathrm{C}= 3.5 (1)$\,K. The temperature dependence of the superfluid density obtained from transverse-field $\mu$SR spectra is best modeled using an isotropic Bardeen-Cooper-Schrieffer type $s$-wave gap symmetry with $2\Delta/k_\mathrm{B}T_\mathrm{C} = 2.84(2)$. However, the zero-field muon-spin relaxation asymmetry reveals the appearance of a spontaneous magnetic field below $T_\mathrm{C}$, indicating that time-reversal symmetry (TRS) is broken in the superconducting state. {Although this behavior is commonly associated with non-unitary or mixed singlet-triplet pairing, our group-theoretical analysis of the Ginzburg-Landau free energy alongside density functional theory calculations indicates that unconventional mechanisms are pretty unlikely. Therefore, we have hypothesized that TRS breaking may occur via a conventional electron-phonon process.}  
   
\end{abstract}
\maketitle

\section{Introduction}
Ternary rare earth-transition metal silicides and germanides have recently been the subject of several studies because of their {wide variety} of structures and phenomenologies ~\cite{Parthe}. The occurrence of superconductivity is commonly found in these compounds, with at least 11 superconducting distinct structural prototypes reported so far~\cite{Braun}. Some of them host $3d$ magnetic species, such as Fe, Co, and Ni, concomitantly with a superconducting ground state~\cite{Braun1980,Braun19,Seqre}. For example, two-gap superconductivity in Lu$_{2}$Fe$_{3}$Si$_{5}$ has been reported from low-temperature specific-heat, penetration depth, and muon spectroscopy measurements~\cite{Nakajima,Gordon,Biswas2011}. The coexistence of charge density waves with magnetic or superconducting order is also reported in $R_{5}T_{4}$Si$_{10}$ ($R = $~rare earth element; $T = $~Co, Ir, Rh, or Os) family~\cite{Yang,Ghosh,Becker,Galli,Kuo,Smaalen}. {These observations highlight the potential to use this class of materials to study different and novel quantum states.}

The highest superconducting critical temperature $\left(T_\mathrm{C}\right)$ reported at ambient pressures is due to Y$_{5}$Os$_{4}$Si$_{10}$~\cite{Hausermann}, with a $T_\mathrm{C}$ of 9.1\,K, followed by Sc$_{5}$Ir$_{4}$Si$_{10}$ ($T_\mathrm{C} = 8.6$\,K), and Sc$_{5}$Rh$_{4}$Si$_{10}$ ($T_\mathrm{C} = 8.4$\,K). The occurrence of superconductivity in Sc$_{5}$Co$_{4}$Si$_{10}$ is particularly noteworthy. Sc and Si are not superconducting at ambient pressure, while Co is a ferromagnetic material with no superconducting state at all. Superconductivity in materials containing Co is relatively unusual. Examples include the binary phase CoSi$_{2}$ with a $T_\mathrm{C} = 1.2$\,K~\cite{Matthias} and Lu$_{3}$Co$_{4}$Ge$_{13}$ single crystals with $T_\mathrm{C}$ below 1.4\,K~\cite{rai2015}. Superconductivity is also reported in layered materials such as NaCo$_x$O$_2\cdot y$H$_2$O at $\sim 5$\,K~\cite{Takada2003} and, more recently, LaCoSi below 4\,K~\cite{He2021}. Additionally, the quasi-one-dimensional carbide Sc$_3$CoC$_4$~\cite{Scherer2010} becomes superconducting at approximately 4.5\,K.

\begin{figure*}[tb]
\centering
\includegraphics[width=0.9\linewidth]{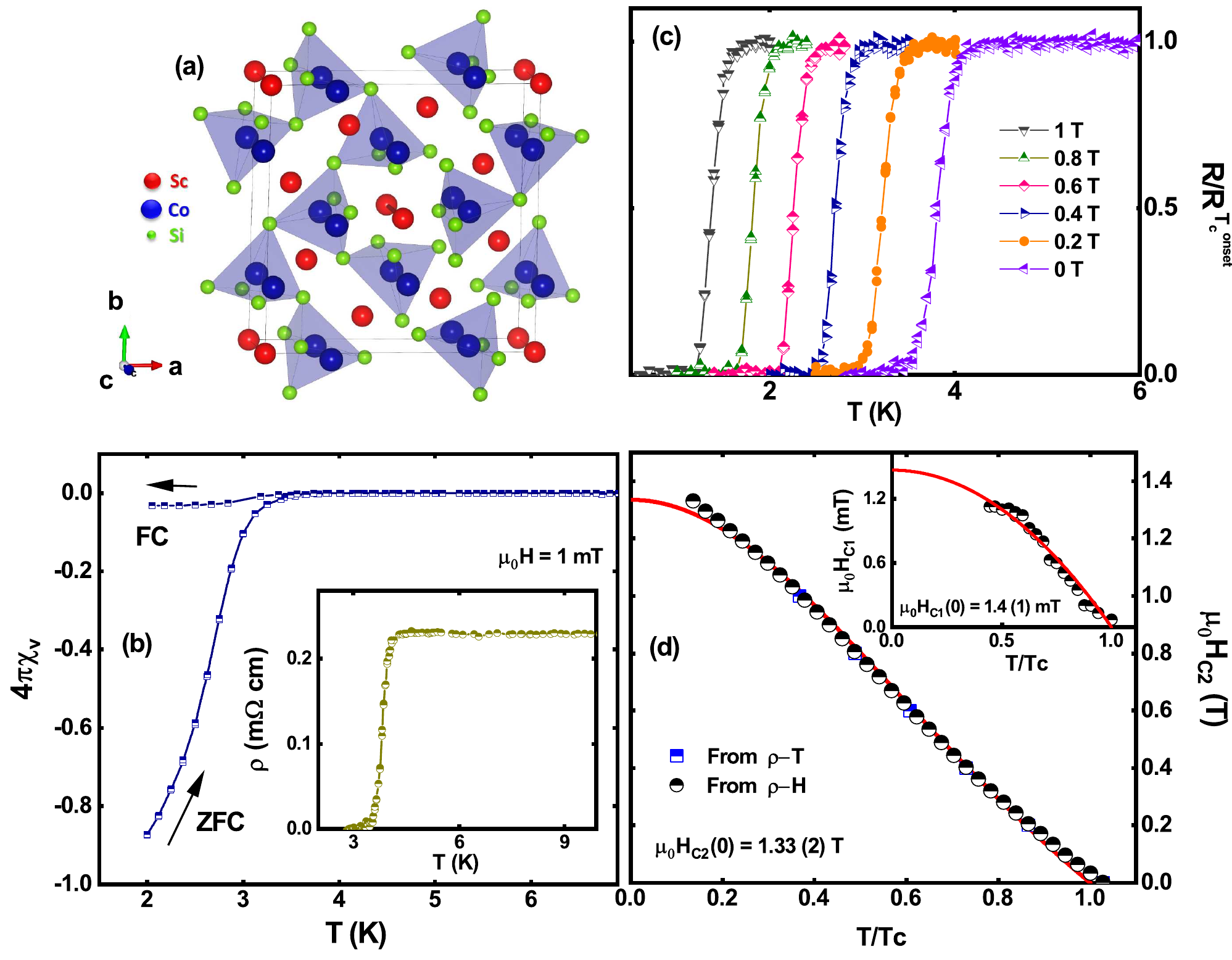}\hfil
\caption{(a) Unit cell of Sc$_{5}$Co$_{4}$Si$_{10}$ viewed along the $c$~axis. The Co (blue) and Si (green) atoms form pentagons and hexagons stacked parallel to the basal plane interspersed with Sc (red) atoms. (b) Temperature dependence of the magnetic susceptibility $\chi\left(T\right)$ for Sc$_{5}$Co$_{4}$Si$_{10}$, collected in zero-field-cooled (ZFC) and field-cooled (FC) modes in an applied field of 1\,mT. The inset shows the resistivity versus temperature for Sc$_{5}$Co$_{4}$Si$_{10}$ in zero applied field. (c) Temperature dependence of the normalized resistivity for Sc$_{5}$Co$_{4}$Si$_{10}$ in different applied magnetic fields. (d) Temperature dependence of the upper critical field $H_{\mathrm{C2}}$ of Sc$_{5}$Co$_{4}$Si$_{10}$ determined using resistivity measurements made as a function of either temperature or applied magnetic field. $H_\mathrm{C2}(0)$ is estimated from the fit shown by a red line using the GL expression $H_\mathrm{C2}(T) = H_\mathrm{C2}(0)\left(1-t^2\right)/\left(1+t^2\right)$. The inset shows the lower critical field $H_{\mathrm{C1}}$ for Sc$_{5}$Co$_{4}$Si$_{10}$ as a function of the reduced temperature $t = T/T{_\mathrm{C}}$ fitted using the GL expression $H_{\mathrm{C1}}(T) = H_{\mathrm{C1}}(0)(1-t^{2}$).}
\label{mtrtcp}
\end{figure*}

The {relatively} high $T_\mathrm{C}$ of Sc$_{5}$Co$_{4}$Si$_{10}$ ($T_\mathrm{C} = 4.9$\,K) indicates that the magnetic character of Co atoms is either suppressed or does not interfere with the electron pairing mechanism~\cite{Braun1980}. Sc$_{5}$Co$_{4}$Si$_{10}$ crystallizes in a tetragonal structure within the centrosymmetric space group $P4/mbm$ (No. 127), with 38 atoms per unit cell~\cite{Parthe,Braun1980,Braun19,Seqre}. There are three nonequivalent Sc sites in the structure, one of which forms a chain-like Sc--Si network along the $c$~axis. One can see from the projection of the Sc$_{5}$Co$_{4}$Si$_{10}$ structure along the $c$~axis [Fig.~\ref{mtrtcp}(a)] that Co and Si atoms give rise to planar networks of pentagons and hexagons stacked parallel to the basal plane and connected along the $c$~axis via Co--Si--Co zigzag chains. The pentagon-hexagon layers are interspersed with layers of Sc~\cite{Braun80}. One interesting feature is that {Co--Co} distance is of the order of 4\,\AA, far larger than in pure Co. This is quite different from what is found in the Mo chalocogenides~\cite{Fischer} and rhodium borides~\cite{Vandenberg}, both well studied ternary systems in which the transition metal atoms form clusters and play an important role in superconductivity. Nakajima~{\it et al.}~\cite{Nakajima2009} reported the low-temperature heat capacity of Sc$_{5}$Ir$_{4}$Si$_{10}$, showing that the data is best described by a two-gap model. Nuclear magnetic resonance measurements on Sc$_{5}$Co$_{4}$Si$_{10}$ suggest that the density of states at the Fermi level is dominated by the Co-$d$ bands derived from the pentagon-hexagon layers {. In contrast}, the $d$ bands of Sc layers, which separate the pentagon--hexagon sheets, are nearly empty~\cite{Koyama,Lue}.

Motivated by these results, we have conducted a systematic study of the superconducting properties of Sc$_{5}$Co$_{4}$Si$_{10}$ {employing} electrical resistivity, magnetization, heat capacity, and muon-spin rotation and relaxation ($\mu$SR) measurements. The thermal, transport, and magnetic measurements indicate a bulk type-II superconducting ground state with $T_\mathrm{C}=3.5(1)$\,K, $\mu_0H_\mathrm{C1}=1.4(1)$\,mT, and $\mu_0H_\mathrm{C2} = 1.33(2)$\,T. The temperature dependence of the superfluid density estimated from transverse-field (TF) $\mu$SR measurements shows the presence of a fully-gapped superconducting order parameter. This finding is also supported by heat capacity measurements at low temperatures. Surprisingly, zero-field (ZF) $\mu$SR measurements {reveal a finite}, spontaneous internal magnetic field within the superconducting state, providing clear evidence for time-reversal symmetry breaking. A first analysis based on {Ginzburg-Landau's theory alongside density functional theory suggests that both non-unitary triplet pairing state and mixed singlet-triplet state are quite unlikely due to the complex, closed topography of the Fermi surface. Therefore, we argue that conventional mechanisms could carry TRS breaking due to the highly symmetric multiband low-energy states.} Sc$_{5}$Co$_{4}$Si$_{10}$, therefore, {joins the shortlist} of unconventional TRS breaking superconducting materials with fully-gapped $s$-wave symmetry.

\section{Methods}

\subsection{Experimental Details}

Polycrystalline Sc$_{5}$Co$_{4}$Si$_{10}$ samples were prepared by arc melting stoichiometric amounts of high purity elemental Sc (99.9\%), Co (99.99\%), and Si (99.9999\%) in an argon atmosphere with a Zr getter. The ingot was flipped and remelted several times to improve the phase homogeneity. The sample was then annealed in a dynamic vacuum of $10^{-6}$\,torr for 18\,days at a temperature of $1050\,^{\circ}$C, followed by a quench to room temperature by switching off the heater power. {The homogeneous single phase nature of the sample with a 5:4:10 composition for Sc:Co:Si was confirmed using a ZEISS GeminiSEM 500, which was used to perform energy-dispersive X-ray spectroscopy (EDX)}. Electrical resistivity measurements were carried out in different applied fields using a Quantum Design Physical Property Measurement System (QD-PPMS). The $dc$-magnetization measurements were performed using a Quantum Design Magnetic Property Measurement System SQUID magnetometer. Heat capacity down to 0.5\,K was measured using a QD-PPMS with a $^3$He insert.

Muon spin rotation and relaxation ($\mu$SR) measurements were carried out using the {MuSR} spectrometer at the ISIS Pulsed Neutron and Muon Source, United Kingdom~\cite{Lee1999,Bhattacharyya1321}. The powdered sample was mounted on a high purity silver plate (99.995\%) using GE varnish, and then covered with silver foil. The sample was cooled to temperatures as low as 400\,mK using a He-3 system. 100\% spin-polarized positive muons ($\mu^{+}$) were implanted into the sample. Each $\mu^{+}$ decays with a mean lifetime of $2.2\,\mu$s, releasing a positron. The time dependence of the polarization of the implanted muons is given by $P_\mathrm{\mu}(t) = G(t)P_\mathrm{\mu}(0)$, where $G(t)$ corresponds to the $\mu^{+}$ spin auto-correlation function. The time-dependent asymmetry $A(t)$, which is proportional to $P_{\mu}(t)$, is given by $A(t) = \left[ {N_{\mathrm{F}}(t) -\alpha N_{\mathrm{B}}(t)}\right]/\left[{N_{\mathrm{F}}(t)+\alpha N_{\mathrm{B}}(t)}\right]$, where $N_{\mathrm{B}}(t)$ and $N_{\mathrm{F}}(t)$ are the number of positrons counted in the backward and forward detectors, respectively, and $\alpha$ is an instrumental calibration constant determined in the normal state in a small (2\,mT) transverse magnetic field. All the $\mu$SR data were analyzed using the WiMDA data analysis program~\cite{Pratt2000}. The TF-$\mu$SR data were collected at different temperatures between 0.4 and 5\,K in a transverse-field of 40\,mT ($> H_\mathrm{C1}(0)$, which is $\approx 1.4(1)$\,mT). The ZF data were collected between 1.3 and 6.2\,K, and active compensation coils were used to reduce any stray magnetic field at the sample position to below $\sim 0.1\,\mu$T.

\subsection{Computational methods}

Calculations were performed using density functional theory (DFT) as implemented in the Vienna \textit{ab initio} simulation package (VASP)~\cite{kresse1993,kresse1996,kresse1996_2}. Spin-polarized and spin-orbit coupling calculations were performed with the projector-augmented wave (PAW) method using a plane-wave basis set of 500\,eV~\cite{blochl1994,kresse1999} using the Perdew-Burke-Ernzerhof (PBE) functional~\cite{perdew1996}. The following electrons were treated explicitly: 3$p^6$ 4$s^2$ 3$d^3$ (Sc), 3$d^8$ 4$s^1$ (Co), and 3$s^2$ 3$p^2$ (Si). Electronic relaxations were carried out until the self-consistent calculation loops reached an energy convergence of less than $10^{-5}$\,eV and structural relaxations were performed until forces on all atoms were below 0.01\,eV/Å. A k-point mesh of $8\times8\times24$ was used to sample the Brillouin zone (BZ).

\begin{figure}[tb]
\centering
\includegraphics[width=0.9\linewidth]{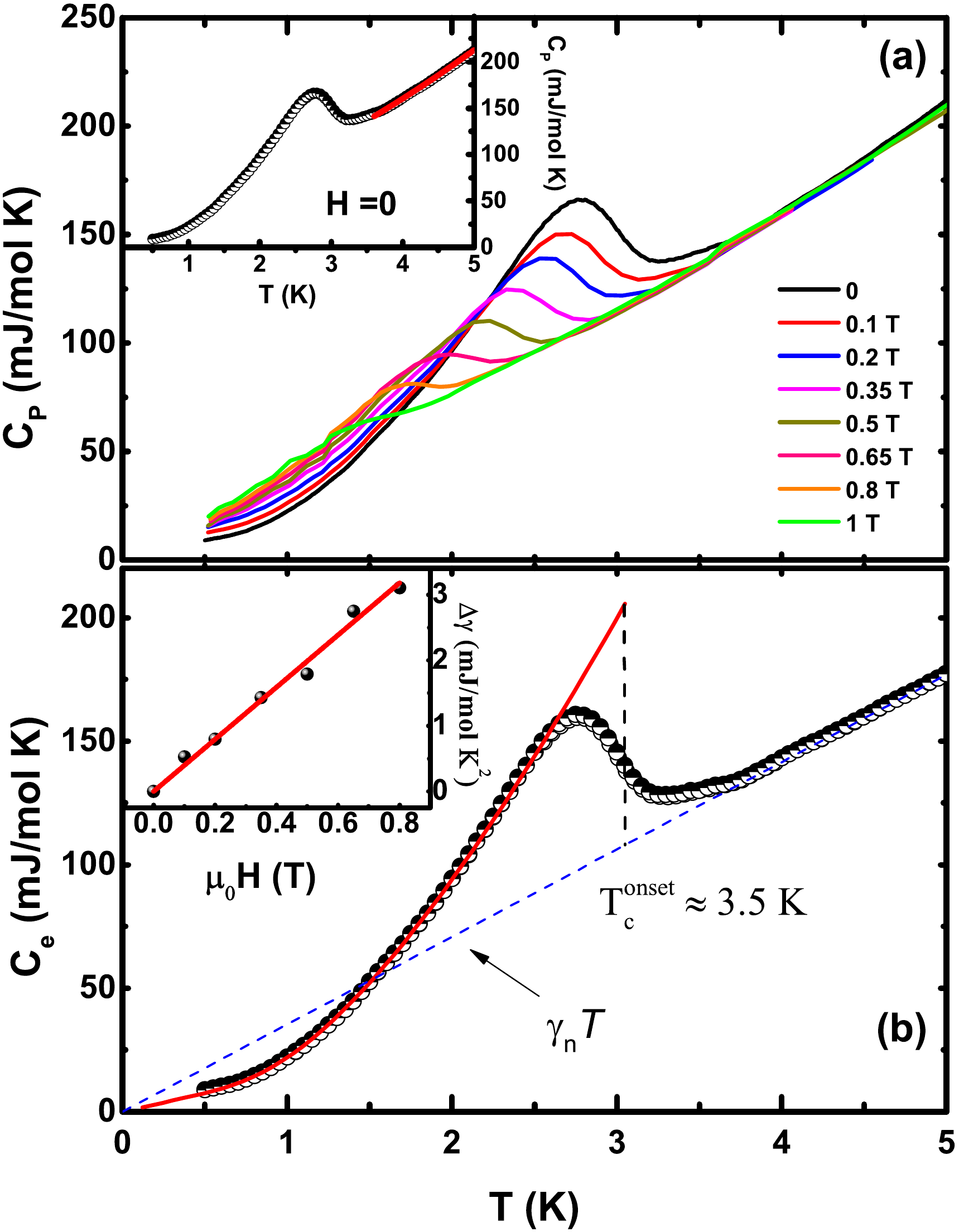}\hfil
\caption{(a) Temperature dependence of heat capacity $C_\mathrm{P}(T)$ for $0.4\leqslant T \leqslant 5$\,K measured in different applied magnetic fields. The inset shows the fit of $C_\mathrm{p}$ in the normal state. (b) Electronic contribution to the zero-field heat capacity, $C_\mathrm{e}$, as a function of temperature. The solid line indicates the fit to $C_\mathrm{e}$ with the isotropic Bardeen-Cooper-Schrieffer expression and the normal state contribution ($\gamma_{ns}T$). The inset shows the field dependence of $\Delta\gamma$. The linear trend follows the behavior expected for conventional BCS-type superconductors. {The blue dashed line represents the extrapolation of normal state electronic heat capacity $\gamma_{n}(T)$.}}
\label{heat}
\end{figure}

\section{Results and Discussion}
\subsection{Physical characterization}

The temperature dependence of the resistivity $\rho\left(T\right)$ of Sc$_{5}$Co$_{4}$Si$_{10}$ in zero applied field and in different applied fields are shown in Fig.~\ref{mtrtcp}(c). There is a sharp superconducting transition at $T_\mathrm{C} = 3.5(1)$\,K. {The high temperature resistivity data exhibits metallic behavior. The estimated residual resistivity ratio  $\rho(220K) /\rho(4 K)$ = 2.04, this value is close to that seen in the isostructural superconductor Lu$_{5}$Ir$_{4}$Si$_{10}$~\cite{Singh}} The temperature dependence of the magnetic susceptibility $\chi\left(T\right)$ of Sc$_{5}$Co$_{4}$Si$_{10}$ in an applied magnetic field of 1\,mT is shown in Fig.~\ref{mtrtcp}(b). $\chi(T)$ reveals a clear signature of superconductivity below the superconducting transition temperature $T_{\mathrm{C}} = 3.5(1)$\,K as well, with a superconducting volume close to 90\% at 2.0\,K, thus confirming the bulk nature of the superconductivity.

To determine the lower critical field $H_{\mathrm{C1}}$, field-dependent magnetization $M\left(H\right)$ curves for Sc$_{5}$Co$_{4}$Si$_{10}$ were collected at various temperatures up to $T_\mathrm{C}$ using a ZFC-protocol. For each temperature, $H_{\mathrm{C1}}\left(T\right)$ was determined as the field where $M\left(H\right)$ first deviates from linearity. $H_\mathrm{C1}$ as a function of the reduced temperature $T/T_{\mathrm{C}}$ is shown in Fig.~\ref{mtrtcp}(d). The temperature variation of $H_\mathrm{C1}$ is reasonably well described by the Ginzburg-Landau (GL) expression $H_{\mathrm{C1}}(T) = H_{\mathrm{C1}}(0)(1-t^{2}$), where $t = T/T_{\mathrm{C}}$, yielding $\mu_0H{_\mathrm{C1}} (0) = 1.4(1)$\,mT.

The upper critical field $H_\mathrm{C2}$ for Sc$_{5}$Co$_{4}$Si$_{10}$ was estimated using the resistivity data collected at different temperatures and applied magnetic fields. The temperature of the midpoint of the resistivity transition is taken as the transition temperature $T_{\mathrm{C}}\left(H\right)$ in the field $H$. Figure~\ref{mtrtcp}(d) shows $H_\mathrm{C2}(T)$, which follows a nearly linear behavior in the temperature range close to $T_{\mathrm{C}}$ that can be fitted with the phenomenological Ginzburg-Landau expression $H_\mathrm{C2}(T) = H_\mathrm{C2}(0)\frac{1-t^2}{1+t^2}$. The fit gives $\mu_0H_{C2} = 1.33(2)$\,T, well below the Pauli paramagnetic limit $1.84T_\mathrm{C}\left[\mathrm{K}\right] = 6.44$\,T~\cite{WHH}.

Heat capacity $C_{\mathrm{P}}$ as a function of temperature for $0.45\leqslant T \leqslant5$\,K is shown in Fig.~\ref{heat}(a) in different applied magnetic fields. The inset in Fig.~\ref{heat}(a) shows the temperature variation of the heat capacity in zero applied {magnetic fields.} A clear signature of the superconducting transition is observed below 3.5\,K in $C_\mathrm{P}(T)$. Above $T_{\mathrm{C}}$, in the normal state, $C_{\mathrm P}\left(T\right)$ was found to be independent of the external magnetic field and can be described using $C_\mathrm{P}(T) = \gamma T + \beta T^{3}$, where $\gamma $ is the electronic Sommerfeld coefficient and $\beta T^{3}$ is the lattice (phonon) contribution to the specific heat, yielding $\gamma = 35.6(5)  $\,mJ mol$^{-1}$K$^{-2}$ and $\beta = 0.276(1) $\,mJ mol$^{-1}$K$^{-4}$. Using the Debye model, the Debye temperature is given by $\Theta_{\mathrm{D}} = (\frac{12\pi^{4}}{5\beta}nR)^{1/3}$, where $R = 8.314$\,Jmol$^{-1}$K$^{-1}$ is the gas constant and $n = 19$ is the number of atoms per formula unit in Sc$_{5}$Co$_{4}$Si$_{10}$. Using this relationship, $\Theta_{\mathrm{D}}$ is estimated to be 511(3)\,K. Figure~\ref{heat}(b) shows the temperature dependence of the electronic specific heat, $C_{\mathrm{e}}\left(T\right)$, obtained by subtracting the phonon contribution from $C_{\mathrm P}\left(T\right)$. $C_{\mathrm{e}}\left(T\right)$ can be used to investigate the superconducting gap symmetry. Using the maximum entropy construction shown in Fig.~\ref{heat}(b) the heat capacity jump at T$_\mathrm{C}$ is $\Delta C_{e}$ = 96\,mJ/mol K$^{2}$, which yield $\Delta C_{e}/\gamma T_\mathrm{C} \approx$ 0.90, which is smaller than the 1.43 expected for weak-coupling BCS superconductors~\cite{BCS}. We analyzed the superconducting state electronic heat capacity within the framework of BCS model~\cite{BCS}. The solid red curve in Fig.~\ref{heat}(b) corresponds to the fit according to $C_{\rm e} = C_{\rm BCS} + \gamma_{\rm ns} T$, where $C_{\rm BCS}$ represents the theoretical weak-coupling fully gapped isotropic s-wave BCS superconductivity corresponding to $\gamma_{\rm s} = 21$~mJ/mol\,K$^2$, and $\gamma_{\rm ns} T$ represents the contribution due to {the} nonsuperconduting fraction of {the} sample. {The nonsuperconducting contribution might have its origin in defects in materials as well as due to the presence of small inhomogeneous/impurity phase(s).} 

\par

The inset of Fig.~\ref{heat}(b) presents the magnetic field dependence of the Sommerfeld coefficient $\Delta\gamma = \gamma(H)-\gamma(0)$ from values of $\gamma$ obtained by extrapolating $C_{\mathrm{e}}\left(T\right)$ to $T \sim$ 0\,K. Interestingly, the linear trend follows the behavior expected for a conventional BCS-type superconductor.

\begin{figure}[t]
\centering
\includegraphics[width=0.9\linewidth]{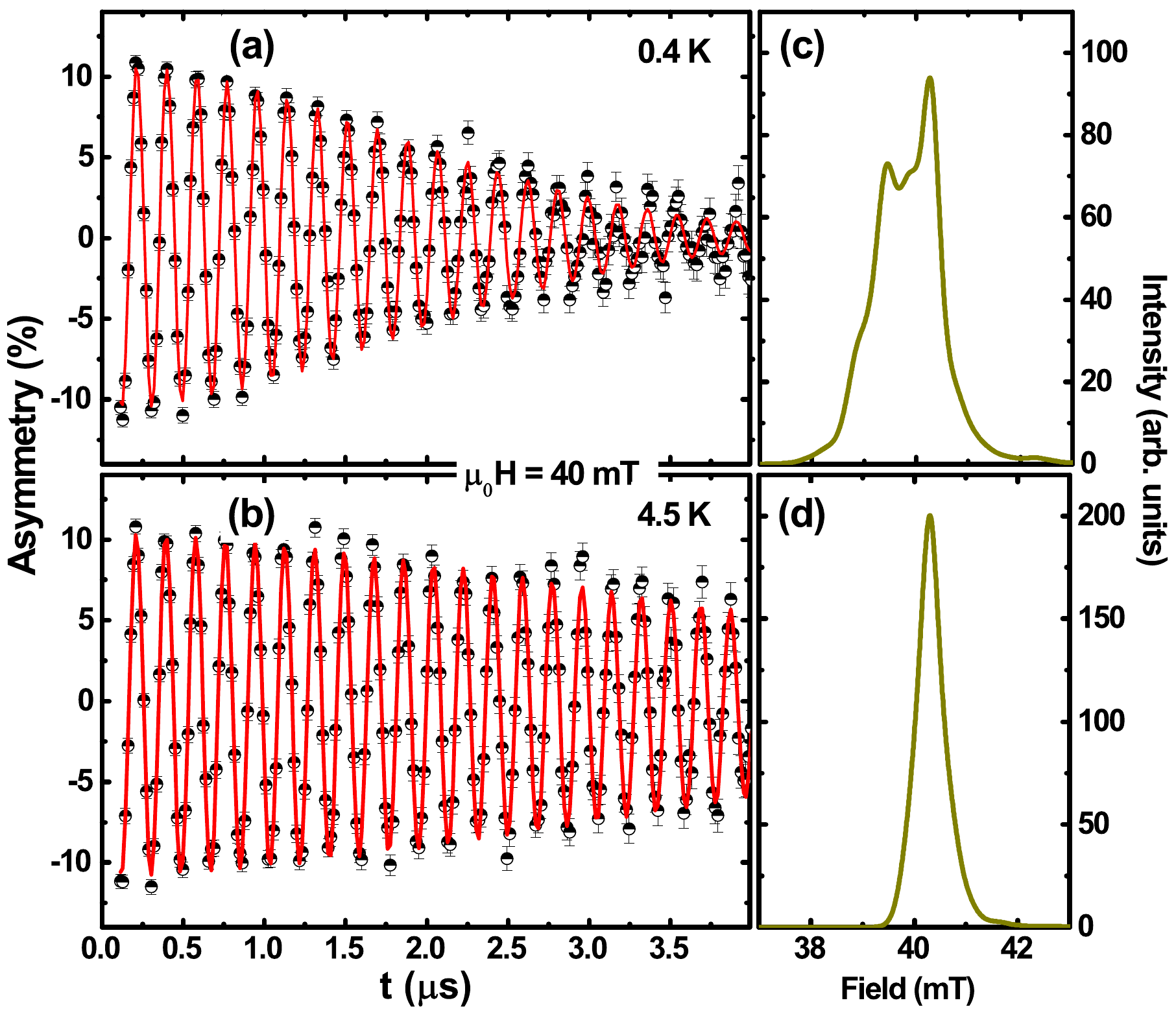}\hfil
\caption{Time evolution of TF-$\mu$SR asymmetry spectra for Sc$_{5}$Co$_{4}$Si$_{10}$ recorded at (a) 0.4\,K (below $T_\mathrm{C}$) and (b) 4.5\,K (above $T_\mathrm{C}$). The solid red lines represent the fit to the data using Eq.~\ref{GTFfunction}. (c) and (d) display the probability of internal field distributions obtained using the  maximum entropy method, below and above $T_\mathrm{C}$, respectively.}
\label{TFasymmetry}
\end{figure}

\begin{figure}[t]
\centering
\includegraphics[width=0.9\linewidth]{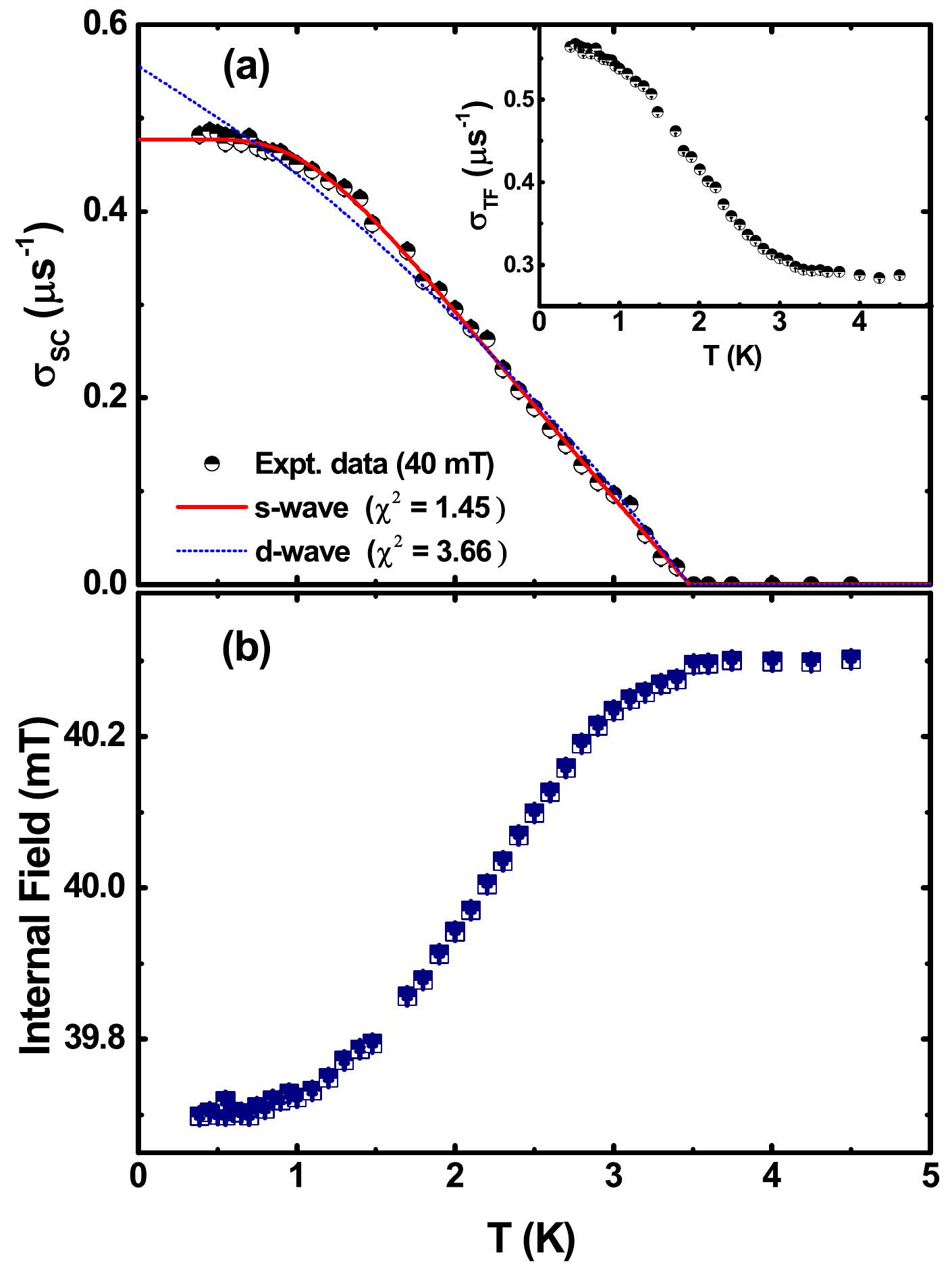}\hfil
\caption{(a) Temperature dependence of the superconducting depolarization rate $\sigma_\mathrm{sc}(T)$ in the presence of an applied magnetic field of 40\,mT. The solid red and dotted blue lines were determined using Eq.~\ref{GTFfunction2} considering both $s$-wave and $d$-wave symmetries, respectively. The inset shows the total muon spin depolarization rate $\sigma_\mathrm{TF}$ as a function of temperature. (b) Temperature dependence of the internal field.}
\label{TFasymmetryfit}
\end{figure} 

\begin{figure*}[t]
\centering
\includegraphics[width=0.9\linewidth]{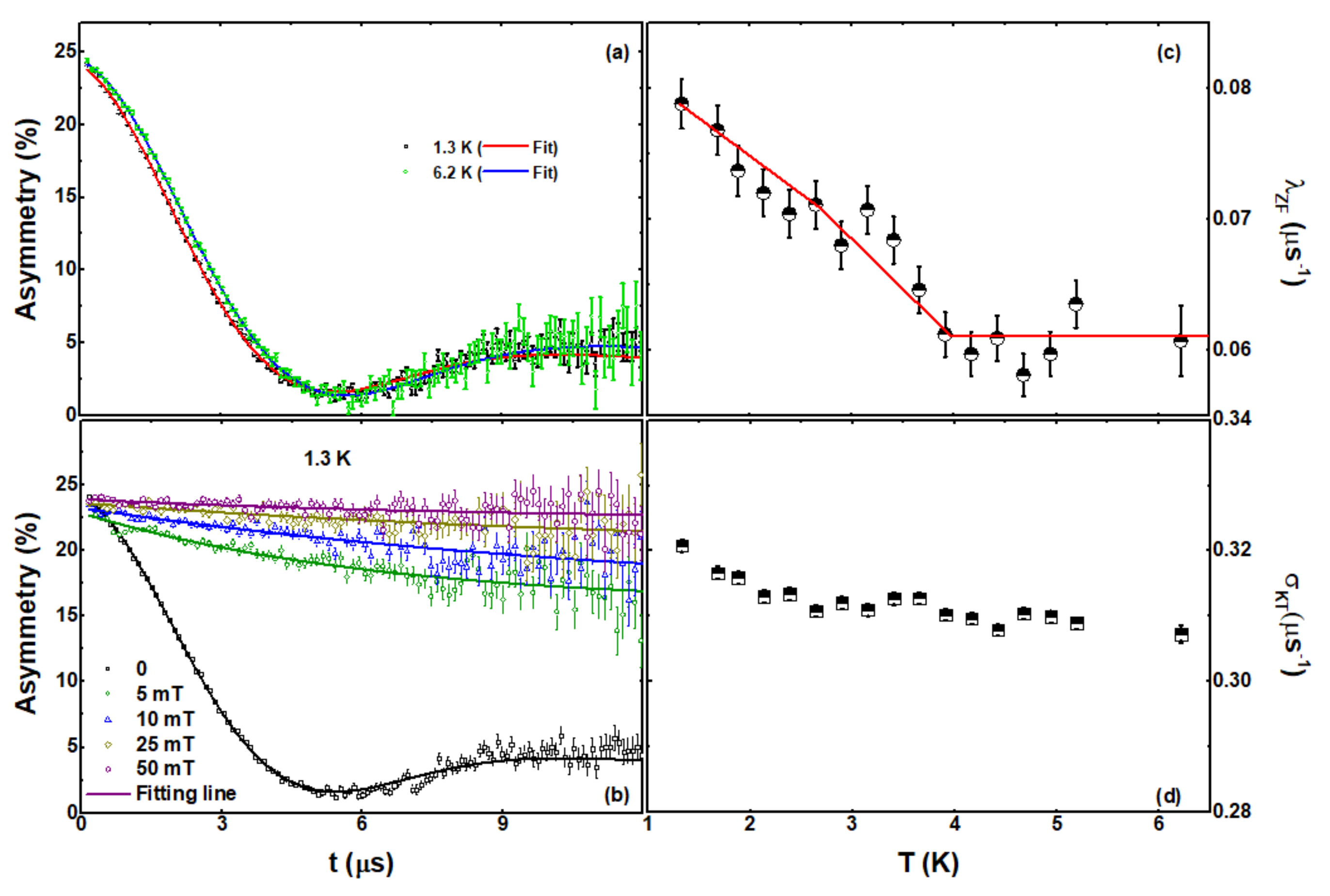}\hfill
\caption{(a) Time evolution of ZF-$\mu$SR asymmetry spectra at 1.3\,K (black squares) and 6.2\,K (green circles). {The red and blue lines are the least square that fit the data using Eq.~\ref{GKTfunction}.} (b) Time dependence of LF-$\mu$SR asymmetry spectra in different applied fields. {The solid lines are the least square that fit the data using Eq.~\ref{GKTfunction}. The line to H = 0 is same as in Fig.(a) and fit lines to other spectra is with $G_\mathrm{KT}$(t)=1 in Eq.~\ref{GKTfunction}. } (c) Temperature dependence of the ZF-relaxation rate, $\sigma_\mathrm{ZF}$. The solid red line is a guide to the eye. (d) Temperature dependence of $\sigma_\mathrm{KT}$, which is almost constant over the temperature range analyzed.}
\label{ZFasymmetry}
\end{figure*}

\subsection{Superconducting gap structure}

{To investigate} the superconducting gap structure in Sc$_{5}$Co$_{4}$Si$_{10}$, we have performed TF-$\mu$SR measurements. Figures~\ref{TFasymmetry}(a) and~\ref{TFasymmetry}(b) display the TF-$\mu$SR asymmetry spectra recorded at temperatures below and above $T_{\mathrm{C}}$, respectively, with an applied magnetic field of 40\,mT, well above $H_\mathrm{C1}(0)$. The corresponding field distributions determined using the maximum entropy method are shown in Figs.~\ref{TFasymmetry}(c) and \ref{TFasymmetry}(d). The presence of a flux-line lattice in the superconducting state results in an inhomogeneous field distribution within the sample, which in turn induces a faster decay in the asymmetry spectra below $T_\mathrm{C}$. The time evolution of the TF-$\mu$SR data at all temperatures above and below $T_{\mathrm{C}}$ is best described by a sinusoidal oscillatory function damped with a Gaussian relaxation arising from muons implanted in the sample and an oscillatory background term from muons in the silver sample holder that do not depolarize~\cite{Bhattacharyyarev, BhattacharyyaThCoC2,CeIr3,Zr5Pt3}:
\begin{equation}\label{GTFfunction}
G_\mathrm{TF}(t) = A_1\cos\left(\omega_\mathrm{1}t+\varphi\right)\exp\left(-\frac{\sigma_\mathrm{TF}^{2}t^{2}}{2}\right)+ A_\mathrm{2}\cos\left(\omega_2t+\varphi\right).
\end{equation}
{Here, $A_1$ is the initial asymmetry of muons in the sample, and $A_2$ is the asymmetry of the background. $\omega_{\mathrm{1}}$ and $\omega_{\mathrm{2}}$ are the muon precession frequencies within the sample, and the sample holder, respectively, and $\varphi$ is an initial phase offset. Finally, $\sigma_\mathrm{TF}$ is the total muon spin relaxation rate and consists of two contributions. One is due to the inhomogeneous field variation across the superconducting vortex lattice, $\sigma_\mathrm{sc}$. The other is a regular state contribution, $\sigma_{\mathrm{n}}$ which is taken to be temperature-independent over the entire temperature range studied and was obtained from spectra measured above $T_\mathrm{C}$.} Fitting reveals $A_1$ is 0.910(1) and $A_2$ is 0.090(1) of the total initial asymmetry, while $\sigma_{\mathrm{n}} = 0.293~\mu s^{-1}$. Using $\sigma_\mathrm{TF}^{2} = \left(\sigma_\mathrm{sc}^{2} + \sigma_\mathrm{n}^2\right)$ we obtain the superconducting contribution $\sigma_{\mathrm{sc}}$. 

The temperature variation of the penetration depth/superfluid density was modeled using~\cite{Prozorov, AdrojaK2Cr3As3,ZrIrSi}
\begin{align}\label{GTFfunction2}
\frac{\sigma_\mathrm{sc}(T)}{\sigma_\mathrm{sc}(0)} &= \frac{\lambda^{-2}\left(T,\Delta_\mathrm{0,i}\right)}{\lambda^{-2}\left(0,\Delta_\mathrm{0,i}\right)}, \nonumber \\
&= 1 + \frac{1}{\pi}\int_{0}^{2\pi}\int_{\Delta\left(T\right)}^{\infty}\left(\frac{\delta f}{\delta E}\right) \frac{EdEd\phi}{\sqrt{E^{2}-\Delta^2(T,\phi})}, 
\end{align}
where $f= \left[1+\exp\left(E/k_\mathrm{B}T\right)\right]^{-1}$ is the Fermi function and $\Delta_\mathrm{i}(T,0) = \Delta_\mathrm{0,i}\delta(T/T_\mathrm{C})\mathrm{g}(\phi)$ is the value of the superconducting gap at 0\,K. The temperature dependence of the superconducting gap is approximated by the relation $\delta(T/T_\mathrm{C}) = \tanh\left[1.82[1.018\left(T_\mathrm{C}/T-1\right)\right]^{0.51}]$, where $\mathrm{g}(\phi$) is the angular dependence of the superconducting gap function. $\mathrm{g}(\phi$) is replaced by 1 for a $s$-wave gap or $\vert\cos(2\phi)\vert$ for a $d$-wave gap with line nodes~\cite{Pang2015, Annet1990}. We find that the data are best modeled  [see Fig.~\ref{TFasymmetryfit}(a)] using a single isotropic $s$-wave gap of 0.43(1)\,meV, which yields a gap to $T_\mathrm{C}$ ratio 2$\Delta/k_\mathrm{B}T_\mathrm{C} = 2.84(2) $, indicating a weak-coupling regime. {This value of $2\Delta/k_{B}T_{C}$ is significantly smaller than the BCS expected value of 3.53 in weak coupling limit. Usually, a lower value of $2\Delta/k_{B}T_{C}$ is caused by the presence of an anisotropic superconducting gap structure in momentum space or due to the presence of a nonsuperconducting fraction in the sample. As our $\mu$SR data reveal an isotropic $s$-wave superconducting gap structure, we attribute this reduction in value of $2\Delta/k_{B}T_{C}$ to the presence of a small nonsuperconducting fraction in the sample. This is further supported by the fact that the superconducting state electronic heat capacity could be described very well by fully gapped ($2\Delta/k_{B}T_{C}$ = 3.53) isotropic $s$-wave BCS superconductivity by adding a contribution from the nonsuperconducting fraction as discussed above [see Fig. \ref{heat}(b)].
} The muon spin depolarization rate associated with the superconducting state, $\sigma_\mathrm{sc}$, is related with the penetration depth via $\sigma_\mathrm{sc}(T) = 0.06091\frac{\gamma_{\mu}\Phi_{0}}{\lambda_{L}^{2}(T)}$~\cite{Brandt}, where $\Phi_0 = 2.609 \times 10^{-15}$\,Wb is the magnetic flux quantum, $\gamma_{\mu} = 2\pi \times 135.5$\,MHz/T is the muon gyromagnetic ratio. {This gives $\lambda_\mathrm{L}(0) = 259(4)$\,nm from the $s$-wave model fit}. The London model provides a direct relation between $\lambda_{L}(T)$ and ($m^{*}/n_\mathrm{s})(\lambda_\mathrm{L}^2=m^{*}c^{2}/4\pi n_\mathrm{s} e^2)$ where m$^{*} = (1+\lambda_{\mathrm{e-ph}})m_{\mathrm{e}}$ is the effective mass in units of the bare electron mass $m_{\mathrm{e}}$, and $n_{\mathrm{s}}$ is the carrier density. $\lambda_{\mathrm{e-ph}}$ is calculated from $\Theta_{D}$ and $T_\mathrm{C}$ using McMillan equation $\lambda_\mathrm{e-ph} = \frac{1.04+\mu^{*}\ln(\Theta_\mathrm{D}/1.45T_\mathrm{C})}{(1-0.62\mu^{*})\ln(\Theta_\mathrm{D}/1.45T_\mathrm{C})-1.04}$. The superconducting carrier density is then estimated to be $n_\mathrm{s} = 2.7(3) \times 10^{26}$\,carriers\,m$^{-3}$ and the effective-mass enhancement $m^{*} = 1.51(2)\,m_\mathrm{e}$. Details of similar calculations can be found in Refs.~\cite{Chia,Amato,McMillan,HfIrSi}. In Fig.~\ref{TFasymmetryfit}(b), we have plotted the internal magnetic field as a function of temperature. As the sample goes through the superconducting state ($T<T_\mathrm{C}$), the internal magnetic field distribution in the mixed state is less than the applied fields, which clearly {demonstrates} the Meissner field expulsion in the superconducting state.

\subsection{Time-reversal symmetry breaking}

Figure~\ref{ZFasymmetry}(a) shows the time evolution of the zero-field muon-spin relaxation asymmetry in Sc$_{5}$Co$_{4}$Si$_{10}$ at temperatures above (6.2~K) and below (1.3~K) $T_\mathrm{C}$. Below $T_\mathrm{C}$, the muon-spin relaxation becomes faster. It should be noted that there is no signature of muon spin precession, \emph{i.e.} oscillations in $A(t)$, that would follow any internal magnetic field produced by an ordering of the electronic moments.

The ZF-$\mu$SR spectra can be well described by a damped Gaussian Kubo-Toyabe (KT) function
\begin{equation}
G_\mathrm{ZF}(t) = A_\mathrm{3}G_\mathrm{KT}(t)e^{-\lambda_\mathrm{ZF}t}+A_\mathrm{bg},
\label{GKTfunction}
\end{equation}
\noindent where
\begin{equation}
G_{\mathrm{KT}}(t) =\left[ \frac{1}{3}+\frac{2}{3}\left(1-\sigma_\mathrm{KT}^{2}t^{2}\right)\exp\left({-\frac{\sigma_{\mathrm{KT}}^2t^2}{2}}\right)\right].
\label{GKTfunction2}
\end{equation}
\noindent $A_{\mathrm{3}}$, $A_{\mathrm{bg}}$, $\lambda_\mathrm{ZF}$, and $\sigma_\mathrm{KT}$ are the initial asymmetry from muons implanted in the sample, the asymmetry arising from muons landing in the silver sample holder, the electronic relaxation rate {originating from any electronic moment present in the sample}, and the Kubo-Toyabe depolarization rate, respectively. $A_{\mathrm{bg}}$ is found to be temperature independent. {The contribution to $\sigma_\mathrm{KT}$ comes from $^{45}$Sc with nuclear spin $I$=7/2 (natural abundance 100\%) and $^{59}$Co, $I$=7/2 (100\%). $^{28}$Si does not have a nuclear spin, $I$=0 and hence it does not contribute to $\sigma_\mathrm{KT}$. A similar value of $\sigma_\mathrm{KT}$ is also obtained for Sc$_{5}$Rh$_{6}$Sn$_{18}$~\cite{Sc5Rh6Sn18}.}

Figure~\ref{ZFasymmetry}(c) shows that the temperature dependence of the electronic relaxation rate, $\lambda_{\mathrm{ZF}}\left(T\right)$, is constant within error above $T_\mathrm{C}$, and then increases below $T_\mathrm{C}$ down to the lowest temperature measured. {Over the temperature range studied, the Kubo-Toyabe relaxation rate $\sigma_\mathrm{KT}(T)$ remains almost constant, as shown in Fig.~\ref{ZFasymmetry}(d) while $\lambda_{ZF}$ increases below $T_\mathrm{C}$, most probably due to a Lorentzian contribution to the distribution of static fields. Later we have shown that a weak longitudinal field is sufficient to decouple the relaxation (Figure~\ref{ZFasymmetry} b), suggesting that the increase in $\lambda_\mathrm{ZF}$ is due to static rather than dynamic fields.} 

The increase in $\lambda_{\mathrm{ZF}}(T)$ indicates the appearance of a finite, spontaneous internal magnetic field with a Lorentzian distribution correlated with the superconductivity. The increase in $\Delta\lambda_{\mathrm{ZF}}$ below $T_\mathrm{C}$ is 0.017\,$\mu$s$^{-1}$, which corresponds to a characteristic field strength $\Delta\lambda_{\mathrm{ZF}}/\gamma_{\mu} = 0.020$\,mT, where $\gamma_{\mu} = 2\pi \times 135.5$\,MHz/T is the muon gyromagnetic ratio. {Assuming the relaxation is due to the static field we can then use the increase in the exponential relaxation rate, to the field strength. Similar estimates have been made elsewhere, for example, for Sr$_{2}$RuO$_{4}$ the estimated characteristics field strength is 0.05 mT~\cite{Luke}, similar calculations have also been performed for UPt$_{3}$~\cite{luke1993muon}. } This observation provides strong evidence that time-reversal symmetry is broken in the superconducting state of Sc$_{5}$Co$_{4}$Si$_{10}$. Similar {small} changes in $\lambda_{ZF}\left(T\right)$ have provided evidence of TRS breaking in superconducting LaNiC$_2$~\cite{LaNiC2}, $A_{5}$Rh$_{6}$Sn$_{18}$ [$A = $~Y, $R$, or Sc]~\cite{Sc5Rh6Sn18,Y5Rh6Sn18,Lu5Rh6Sn18}, \mbox{SrPtAs}~\cite{SrPtAs}, the binary lanthanides ${\text{La}_{7}}{\text{(Ir, Rh, Pd)}}_{3}$~\cite{Barker2015La7Ir3, Singh2018La7Rh3, Mayoh2021La7Pd3}, the recently discovered ${\text{Zr}}_{3}{\text{Ir}}$~\cite{Sajilesh2019Zr3Ir, Shang2020Zr3Ir}, and, but more controversially, in the layered perovskite ${\text{Sr}}_{2}{\text{RuO}}_{4}$~\cite{Luke, Luke2000Sr2RuO4, Xia2006Sr2RuO4, Kashiwaya2019Sr2RuO4, Grinenko2019Sr2RuO4}. Theoretical analysis has argued that TRS broken state occurs below  $T_\mathrm{C}$ within the BCS formalism in multiband superconductors~\cite{wilson2013}. On the other hand, the TRS breaking has also been proposed theoretically for fully-gap  BCS superconductors based on the loop supercurrent state, as proposed for multi-orbital systems with complex crystal structures~\cite{ghosh2021, ghosh2020}.   

A longitudinal magnetic field of 25\,mT [Fig. \ref{ZFasymmetry}(b)] removes any relaxation due to the spontaneous fields and is sufficient to fully decouple the muons from this relaxation channel. This indicates that the associated magnetic fields are static or quasistatic on the time scale of the muon precession. These observations further support the broken TRS in the superconducting ground state of Sc$_{5}$Co$_{4}$Si$_{10}$. 

\subsection{Theoretical calculations}

{To bridge} the gap between our experimental results suggesting a nodeless, $s$-wave gap structure with broken time-reversal symmetry and a superconducting order parameter coherently underlying such phenomenology, we have used the group theoretical formulation of the Ginzburg-Landau theory alongside first-principles electronic-structure calculations and isotropic single-particle Bogoliubov-de Gennes model to constrain the possible symmetries allowed for the Sc$_5$Co$_4$Si$_{10}$ superconducting order parameter, as explained below.

Considering a continuous gauge-symmetry breaking phase transition, it is possible to fully determine the simplest allowed  symmetries of the superconducting momentum-dependent pairing potential $\hat{\Delta}(\mathbf{k})$ close to the critical temperature by the $\mathcal{D}$-dimensional irreducible representations (IRs) of the crystalline space group \cite{ghosh2020}. For the specific case of the tetragonal space group $P4/mbm$ with point group symmetry $D_{4h}$, there are 8 one-dimensional IRs and 2 two-dimensional IRs available. Based on this, one examine all the possible singlet and triplet order parameters which might occur in Sc$_5$Co$_4$Si$_{10}$ assuming a non-negligible spin-orbit coupling. {Below, we assume a TRS breaking phase, as suggested by muon-spin spectroscopy measurements.} 

\begin{figure*}[t]
\includegraphics[width=\linewidth]{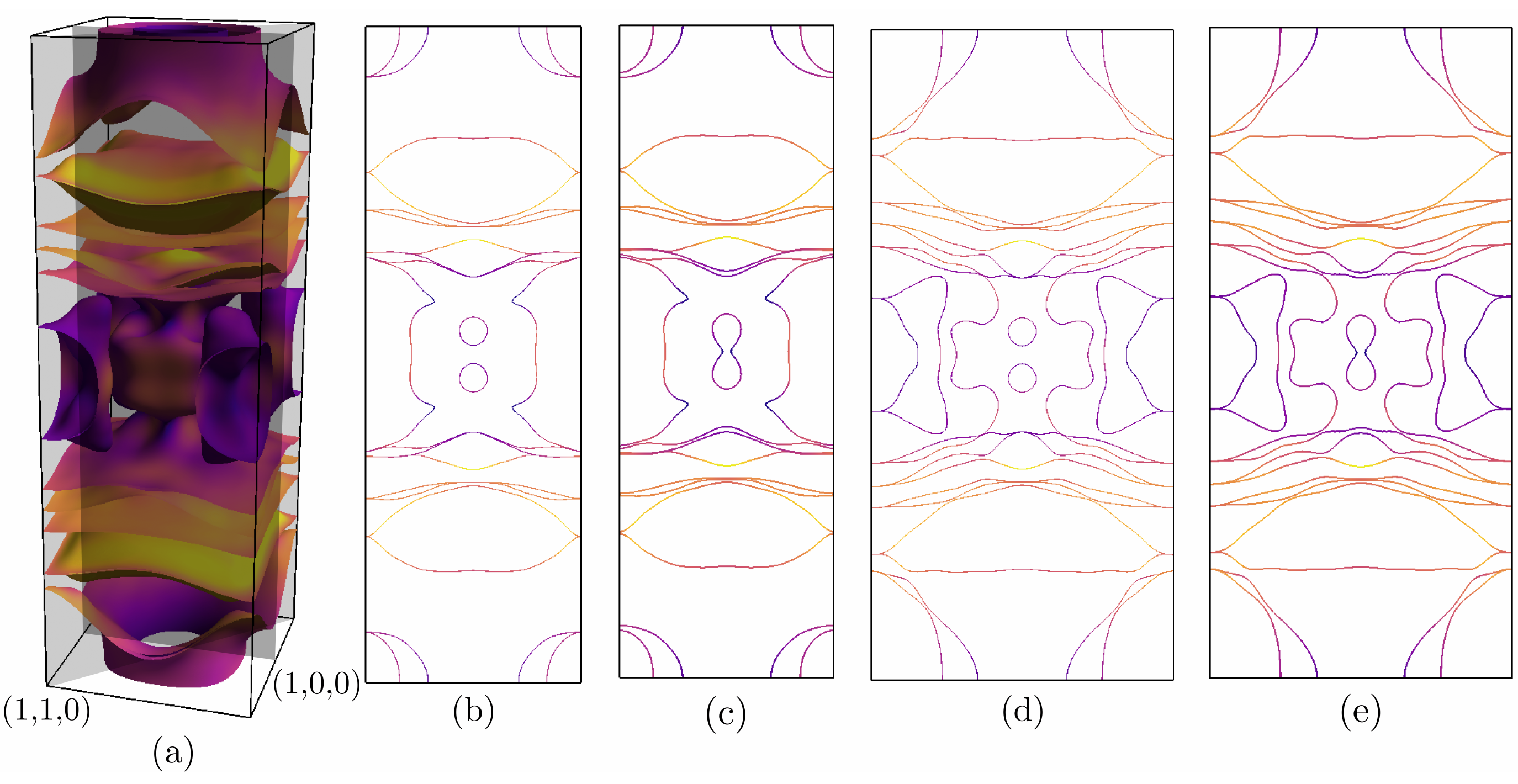}
\caption{Spin-polarized Fermi surface (FS) projected onto the {effective quasiparticle velocity} of Sc$_5$Co$_4$Si$_{10}$. (a) FS representation at the first Brillouin zone. (b--e) Fermi surface sections in the (1,0,0) plane, (b) without and (c) with SOC effects, and in the (1,1,0) plane, (d) without and (e) with SOC effects.}
\label{fig:fermi}
\end{figure*}

In the case of singlet states there are five representations of $SO(3)\times D_{4h}$, namely $^{1}A_{1g}$, $^{1}A_{2g}$, $^{1}B_{1g}$, $^{1}B_{2g}$ and $^{1}E_{g}$, four of which are one-dimensional with a single complex component, and just one, $^{1}E_{g}$, is two-dimensional with two complex components and three types of solution~\cite{Annet1990}. However, it turns out that TRS breaking superconducting ground states require that two or more components of their pairing potential acquire different complex phases with $\mathcal{D}>1$. These conditions stem from the fact that TRS breaking states under the TRS operation should give rise to a new state {with symmetrically non-related order} parameter~\cite{ghosh2020}. As such, among the seven possible singlet states, just $^{1}E_g(c)$ satisfies the conditions above. In this case, by minimizing the Ginzburg-Landau free energy up to {quartic terms}, the broken-time reversal singlet has a $(1,i)$-type ground state with the simplest gap function compatible with its $SO(3)\times D_4(E)\times i$ residual symmetry group being \cite{Annet1990}
\begin{align}
	\hat{\Delta}(\mathbf{k}) = i\left[X(\mathbf{k})+iY(\mathbf{k})\right]Z(\mathbf{k})\hat{\sigma}_y,
\end{align} 
where $X(\mathbf{k})$, $Y(\mathbf{k})$ and $Z(\mathbf{k})$ are the basis functions which transform as $k_x$, $k_y$, and $k_z$, respectively, under point group operations, and $\hat{\sigma}_y$ is the second Pauli matrix. The gap function of the quasiparticle spectrum is then obtained by diagonalising the Bogoliubov-de Gennes Hamiltonian \cite{Lu5Rh6Sn18}
\begin{align}
\mathcal{H}_{BdG} = 
\begin{pmatrix}
	 \hat{h}(\mathbf{k}) & \hat{\Delta}(\mathbf{k})\\
	 \hat{\Delta}^{\dagger}(\mathbf{k}) & -\hat{h}^{T}(-\mathbf{k})\\
\end{pmatrix},
\end{align}
\noindent where $\hat{h}(\mathbf{k})$ is the single-particle Hamiltonian. Thus, for the sake of simplicity, considering the isotropic single-electron dispersion relation, it follows \cite{Lu5Rh6Sn18}
\begin{align}
	\Delta(\mathbf{k}) \propto \left| k_z\right|\sqrt{k_x^2 + k_y^2},
\end{align}   
for the singlet state, yielding a gap structure with line nodes at $k_z = 0$ and two-point nodes at the north ($\theta=0$) and south ($\theta=\pi$) poles of the Fermi surface. 

{In the case of triplet states,} there are also five different IRs of the corresponding double group: the four one-dimensional triplets $A_{1u}$, $A_{2u}$, $B_{1u}$, $B_{2u}$, each with a single complex order parameter, and a two-dimensional representation $E_{u}$, with a complex two-component order parameter and three types of solution~\cite{Annet1990}. Using the same argument as above, just the $E_{u}(c)$ representation with a ground state $(1,i)$ and a $D_4(E)\times i(E)$ residual symmetry group breaks TRS, in which the simplest gap function assumes the general form~\cite{Annet1990}
\begin{align}
	\hat{\Delta}(\mathbf{k}) = i\left[\left(AZ(\mathbf{k}),iAZ(\mathbf{k}),BX(\mathbf{k}) + iBY(\mathbf{k})\right)\cdot\hat{\mathbf{\sigma}}\right]\hat{\sigma}_y,
\end{align} 
where the undetermined coefficients $A$ and $B$ depend on the Fermi surface topography and electronic correlations.

Following the same steps as above, the simplest gap function of the isotropic single-particle  Bogoliubov-de Gennes quasiparticle spectrum for the non-unitary triplet pairing state is given by \cite{Lu5Rh6Sn18}
\begin{align}
\Delta(\mathbf{k}) \propto \left|A\left|k_z\right| - \sqrt{A^2k_z^2 + B^2(k_x^2 + k_y^2)}\right|,
\end{align}
\noindent resulting in gap structures with two point nodes at both the north and south Fermi surface poles. 

In both cases, if the Fermi surface is absent in the regions where the superfluid density is dominated by the low-energy nodal excitations, the superconducting gap would remain fully open over the entire Fermi surface, resembling a nodeless $s$-wave gap structure. This argument is widely used elsewhere~\cite{mackenzie2003,Lu5Rh6Sn18,Y5Rh6Sn18,Shang2020Zr3Ir} to overcome the apparent incompatibility of {TRS breaking} in conventional systems {. However, even in those cases, we believe that} this scenario is quite unlikely since the topography {of the Fermi surface} in real materials is generally very complex and unlikely to meet such conditions. 

{For instance, 12 distinct low-energy bands cross the Sc$_5$Co$_4$Si$_{10}$ Fermi level, resulting in highly complex topography with a three-dimensional character (see Figure \ref{fig:fermi}). Despite the small influence of SOC in low-energy states, several degenerate points are gapped when including the relativistic effects explicitly. The two ball-shaped $\Gamma$-centered pockets have their topography heavily modified by spin-orbit interactions, giving rise to a single lobulated molten spheroid. However, the Dirac-type degeneracies at high-symmetry points are protected against SOC by the double point-group symmetries~\cite{young2012,yang2014,yang2015,ferreira2021}. There are two closed pockets crossing the geometrical poles and several closed electrons- and hole-pockets at the equator along $k_z$ direction, as shown in Fig.\,\ref{fig:fermi}. Therefore, if any, the detection of point nodes would be favored since low-energy states are present in those regions where they are symmetrically allowed.} 
{Recently, Huddart {\it et al.} \textcite{huddart2021} conducted a systematic investigation of muon-stopping sites in TRS breaking superconductors. They have demonstrated that muons lead to self-limited and small high-energy localized perturbations in the electronic structure. Therefore, the observation of time-reversal symmetry breaking superconducting states is unlikely to originate from muon-induced perturbations, including in the present case.}

The conventional electron-phonon mechanism, however, is believed to be enough to give rise to exotic low-symmetry order parameters {in particular conditions}~\cite{agterberg1999}. Since the interband and intraband electron scattering channel are determined by symmetry, it is argued that there is no special need for the resulting superconducting order parameter to preserve the $\mathcal{G}\otimes \mathcal{T}$ symmetry, where $\mathcal{G}$ is the crystal point group and $\mathcal{T}$ is the time-reversal operation. This effect can appear due to the competition between phonon and Coulomb interactions of highly symmetric systems containing multiple electron- and hole-pockets centered at high-symmetry points~\cite{agterberg1999}. Several exotic superconducting compounds have been analyzed recently within this theoretical picture~\cite{mao2003,kreisel2013,singh2017,huang2018,singh2018,shang2018}. {In those cases, it is expected multidimensional gaps of equal magnitude} (that is, hard to differentiate) on each of the Fermi surface sheets within the quasiparticle spectrum. Our calculations show that the Fermi surface topography of Sc$_5$Co$_4$Si$_{10}$ may allow, therefore, a fully-gapped superconducting state with TRS breaking due to conventional mechanisms.  

In some particular cases, an unconventional singlet-triplet pairing can look like a nodeless $s$-wave pairing if the magnitude of the two superconducting gaps is very similar~\cite{Mayoh2021La7Pd3}. Therefore, {additional hypotheses}, including the loop supercurrent state~\cite{ghosh2021}, cannot be ruled out {entirely} in the present case. We hope that the work presented here will motivate further theoretical and experimental studies addressing {the essential} and still open questions on the nature of the pairing mechanisms {in superconductors} that exhibit TRS breaking in what appear to be fully gapped systems.

\section{Summary}

In summary, we have performed a systematic study of the superconducting properties of Sc$_{5}$Co$_{4}$Si$_{10}$ using low-temperature resistivity, magnetization, heat capacity, and muon spin rotation and relaxation measurements. Our measurements show that Sc$_{5}$Co$_{4}$Si$_{10}$ is a type-II superconductor with $T_\mathrm{C} = 3.5(1)$\,K, $\mu_0H_\mathrm{C1} = 1.4(1)$\,mT, and $\mu_0H_\mathrm{C2} = 1.33(2)$\,T. The temperature dependence of the superfluid density obtained from transverse-field $\mu$SR measurements is best modeled assuming an isotropic, single $s$-wave gap structure with 2$\Delta/k_\mathrm{B}T_\mathrm{C} = 2.84(2)$. These findings are supported by the heat capacity measurements. {In addition, the appearance of spontaneous magnetic fields in the ZF-$\mu$SR spectra provides strong evidence for TRS breaking}, suggesting Sc$_{5}$Co$_{4}$Si$_{10}$ may exhibit unconventional pairing, despite the observed full-gapped state. Our theoretical assessment of the symmetry allowed pairing states and the particular features of the Sc$_{5}$Co$_{4}$Si$_{10}$ Fermi surface demonstrate that non-unitary or mixed singlet-triplet order parameters are unlikely, due to the presence of low-energy states in regions where gapless, nodal excitations would be expected. This analysis suggests that conventional mechanisms may be responsible for TRS breaking in Sc$_{5}$Co$_{4}$Si$_{10}$. {Our work paves the way for further studies on this large class of superconductors. Furthermore, it highlights the importance of Sc$_{5}$Co$_{4}$Si$_{10}$ in building a complete understanding of why some materials with solid evidence for TRS breaking simultaneously exhibit behavior expected from more conventional superconductors.}

\begin{acknowledgments}
AB would like to thank Science \& Engineering Research Board for the CRG Research Grant (CRG/2020/000698). DTA is grateful to the EPSRC UK for funding (Grant Ref: EP/W00562X/1), the JSPS for an invitation fellowship, and the Royal Society of London for the International Exchange funding and Newton Advanced Fellowship between UK-China. KP would like to acknowledge the DST India for an Inspire Fellowship (IF170620). This study was financed in part by the Coordenação de Aperfeiçoamento de Pessoal de Nível Superior – Brasil (CAPES) – Finance Code 001. We gratefully acknowledge the S\~ao Paulo Research Foundation (FAPESP) under Grants 2019/05005-7 and 2020/08258-0. The research was partially carried out using high-performance computing resources made available by the Superintend\^encia de Tecnologia da Informa\c c\~ao (STI), Universidade de S\~ao Paulo. The authors also acknowledge the National Laboratory for Scientific Computing (LNCC/MCTI, Brazil) for providing HPC resources of the SDumont supercomputer, which have contributed to the research results reported within this paper. Finally, we would like to thank the ISIS Facility (DOI: 10.5286/ISIS.E.RB1810871) for $\mu$SR beamtime.

\end{acknowledgments}

\bibliographystyle{apsrev4-2}
\bibliography{refs}

\begin{thebibliography}{84}%
\makeatletter
\providecommand \@ifxundefined [1]{%
 \@ifx{#1\undefined}
}%
\providecommand \@ifnum [1]{%
 \ifnum #1\expandafter \@firstoftwo
 \else \expandafter \@secondoftwo
 \fi
}%
\providecommand \@ifx [1]{%
 \ifx #1\expandafter \@firstoftwo
 \else \expandafter \@secondoftwo
 \fi
}%
\providecommand \natexlab [1]{#1}%
\providecommand \enquote  [1]{``#1''}%
\providecommand \bibnamefont  [1]{#1}%
\providecommand \bibfnamefont [1]{#1}%
\providecommand \citenamefont [1]{#1}%
\providecommand \href@noop [0]{\@secondoftwo}%
\providecommand \href [0]{\begingroup \@sanitize@url \@href}%
\providecommand \@href[1]{\@@startlink{#1}\@@href}%
\providecommand \@@href[1]{\endgroup#1\@@endlink}%
\providecommand \@sanitize@url [0]{\catcode `\\12\catcode `\$12\catcode
  `\&12\catcode `\#12\catcode `\^12\catcode `\_12\catcode `\%12\relax}%
\providecommand \@@startlink[1]{}%
\providecommand \@@endlink[0]{}%
\providecommand \url  [0]{\begingroup\@sanitize@url \@url }%
\providecommand \@url [1]{\endgroup\@href {#1}{\urlprefix }}%
\providecommand \urlprefix  [0]{URL }%
\providecommand \Eprint [0]{\href }%
\providecommand \doibase [0]{https://doi.org/}%
\providecommand \selectlanguage [0]{\@gobble}%
\providecommand \bibinfo  [0]{\@secondoftwo}%
\providecommand \bibfield  [0]{\@secondoftwo}%
\providecommand \translation [1]{[#1]}%
\providecommand \BibitemOpen [0]{}%
\providecommand \bibitemStop [0]{}%
\providecommand \bibitemNoStop [0]{.\EOS\space}%
\providecommand \EOS [0]{\spacefactor3000\relax}%
\providecommand \BibitemShut  [1]{\csname bibitem#1\endcsname}%
\let\auto@bib@innerbib\@empty
\bibitem [{\citenamefont {Parth\'{e}}\ and\ \citenamefont
  {Chabot}(1984)}]{Parthe}%
  \BibitemOpen
  \bibfield  {author} {\bibinfo {author} {\bibfnamefont {E.}~\bibnamefont
  {Parth\'{e}}}\ and\ \bibinfo {author} {\bibfnamefont {B.}~\bibnamefont
  {Chabot}},\ }in\ \href@noop {} {\emph {\bibinfo {booktitle} {Handbook on the
  Physics and Chemistry of Rare Earths}}},\ Vol.~\bibinfo {volume} {6},\
  \bibinfo {editor} {edited by\ \bibinfo {editor} {\bibfnamefont {K.~A.}\
  \bibnamefont {Gschneidner}, \bibfnamefont {Jr.}}\ and\ \bibinfo {editor}
  {\bibfnamefont {L.}~\bibnamefont {Eyring}}}\ (\bibinfo  {publisher} {Elsevier
  Science, North-Holland},\ \bibinfo {address} {Netherlands},\ \bibinfo {year}
  {1984})\ pp.\ \bibinfo {pages} {113--334}\BibitemShut {NoStop}%
\bibitem [{\citenamefont {Braun}(1984)}]{Braun}%
  \BibitemOpen
  \bibfield  {author} {\bibinfo {author} {\bibfnamefont {H.~F.}\ \bibnamefont
  {Braun}},\ }\href {https://doi.org/10.1016/0022-5088(84)90057-2} {\bibfield
  {journal} {\bibinfo  {journal} {J. Less Common. Met.}\ }\textbf {\bibinfo
  {volume} {100}},\ \bibinfo {pages} {105} (\bibinfo {year}
  {1984})}\BibitemShut {NoStop}%
\bibitem [{\citenamefont {Braun}\ and\ \citenamefont
  {Segre}(1980)}]{Braun1980}%
  \BibitemOpen
  \bibfield  {author} {\bibinfo {author} {\bibfnamefont {H.~F.}\ \bibnamefont
  {Braun}}\ and\ \bibinfo {author} {\bibfnamefont {C.~U.}\ \bibnamefont
  {Segre}},\ }\href {https://doi.org/10.1016/0038-1098(80)91065-0} {\bibfield
  {journal} {\bibinfo  {journal} {Solid State Commun.}\ }\textbf {\bibinfo
  {volume} {35}},\ \bibinfo {pages} {735} (\bibinfo {year} {1980})}\BibitemShut
  {NoStop}%
\bibitem [{\citenamefont {Braun}(1980)}]{Braun19}%
  \BibitemOpen
  \bibfield  {author} {\bibinfo {author} {\bibfnamefont {H.~F.}\ \bibnamefont
  {Braun}},\ }\href {https://doi.org/10.1016/0375-9601(80)90849-X} {\bibfield
  {journal} {\bibinfo  {journal} {Phys. Lett. A}\ }\textbf {\bibinfo {volume}
  {75}},\ \bibinfo {pages} {386} (\bibinfo {year} {1980})}\BibitemShut
  {NoStop}%
\bibitem [{\citenamefont {Segre}\ and\ \citenamefont {Braun}(1981)}]{Seqre}%
  \BibitemOpen
  \bibfield  {author} {\bibinfo {author} {\bibfnamefont {C.~U.}\ \bibnamefont
  {Segre}}\ and\ \bibinfo {author} {\bibfnamefont {H.~F.}\ \bibnamefont
  {Braun}},\ }\href {https://doi.org/10.1016/0375-9601(81)90334-0} {\bibfield
  {journal} {\bibinfo  {journal} {Phys. Lett. A}\ }\textbf {\bibinfo {volume}
  {85}},\ \bibinfo {pages} {372} (\bibinfo {year} {1981})}\BibitemShut
  {NoStop}%
\bibitem [{\citenamefont {Nakajima}\ \emph {et~al.}(2008)\citenamefont
  {Nakajima}, \citenamefont {Nakagawa}, \citenamefont {Tamegai},\ and\
  \citenamefont {Harima}}]{Nakajima}%
  \BibitemOpen
  \bibfield  {author} {\bibinfo {author} {\bibfnamefont {Y.}~\bibnamefont
  {Nakajima}}, \bibinfo {author} {\bibfnamefont {T.}~\bibnamefont {Nakagawa}},
  \bibinfo {author} {\bibfnamefont {T.}~\bibnamefont {Tamegai}},\ and\ \bibinfo
  {author} {\bibfnamefont {H.}~\bibnamefont {Harima}},\ }\href
  {https://doi.org/10.1103/PhysRevLett.100.157001} {\bibfield  {journal}
  {\bibinfo  {journal} {Phys. Rev. Lett.}\ }\textbf {\bibinfo {volume} {100}},\
  \bibinfo {pages} {157001} (\bibinfo {year} {2008})}\BibitemShut {NoStop}%
\bibitem [{\citenamefont {Gordon}\ \emph {et~al.}(2008)\citenamefont {Gordon},
  \citenamefont {Vannette}, \citenamefont {Martin}, \citenamefont {Nakajima},
  \citenamefont {Tamegai},\ and\ \citenamefont {Prozorov}}]{Gordon}%
  \BibitemOpen
  \bibfield  {author} {\bibinfo {author} {\bibfnamefont {R.~T.}\ \bibnamefont
  {Gordon}}, \bibinfo {author} {\bibfnamefont {M.~D.}\ \bibnamefont
  {Vannette}}, \bibinfo {author} {\bibfnamefont {C.}~\bibnamefont {Martin}},
  \bibinfo {author} {\bibfnamefont {Y.}~\bibnamefont {Nakajima}}, \bibinfo
  {author} {\bibfnamefont {T.}~\bibnamefont {Tamegai}},\ and\ \bibinfo {author}
  {\bibfnamefont {R.}~\bibnamefont {Prozorov}},\ }\href
  {https://doi.org/10.1103/PhysRevB.78.024514} {\bibfield  {journal} {\bibinfo
  {journal} {Phys. Rev. B}\ }\textbf {\bibinfo {volume} {78}},\ \bibinfo
  {pages} {024514} (\bibinfo {year} {2008})}\BibitemShut {NoStop}%
\bibitem [{\citenamefont {Biswas}\ \emph {et~al.}(2011)\citenamefont {Biswas},
  \citenamefont {Balakrishnan}, \citenamefont {Paul}, \citenamefont {Lees},\
  and\ \citenamefont {Hillier}}]{Biswas2011}%
  \BibitemOpen
  \bibfield  {author} {\bibinfo {author} {\bibfnamefont {P.~K.}\ \bibnamefont
  {Biswas}}, \bibinfo {author} {\bibfnamefont {G.}~\bibnamefont
  {Balakrishnan}}, \bibinfo {author} {\bibfnamefont {D.~M.}\ \bibnamefont
  {Paul}}, \bibinfo {author} {\bibfnamefont {M.~R.}\ \bibnamefont {Lees}},\
  and\ \bibinfo {author} {\bibfnamefont {A.~D.}\ \bibnamefont {Hillier}},\
  }\href {https://doi.org/10.1103/PhysRevB.83.054517} {\bibfield  {journal}
  {\bibinfo  {journal} {Phys. Rev. B}\ }\textbf {\bibinfo {volume} {83}},\
  \bibinfo {pages} {054517} (\bibinfo {year} {2011})}\BibitemShut {NoStop}%
\bibitem [{\citenamefont {Yang}\ \emph {et~al.}(1991)\citenamefont {Yang},
  \citenamefont {Klavins},\ and\ \citenamefont {Shelton}}]{Yang}%
  \BibitemOpen
  \bibfield  {author} {\bibinfo {author} {\bibfnamefont {H.~D.}\ \bibnamefont
  {Yang}}, \bibinfo {author} {\bibfnamefont {P.}~\bibnamefont {Klavins}},\ and\
  \bibinfo {author} {\bibfnamefont {R.~N.}\ \bibnamefont {Shelton}},\ }\href
  {https://doi.org/10.1103/PhysRevB.43.7688} {\bibfield  {journal} {\bibinfo
  {journal} {Phys. Rev. B}\ }\textbf {\bibinfo {volume} {43}},\ \bibinfo
  {pages} {7688} (\bibinfo {year} {1991})}\BibitemShut {NoStop}%
\bibitem [{\citenamefont {Ghosh}\ \emph {et~al.}(1993)\citenamefont {Ghosh},
  \citenamefont {Ramakrishnan},\ and\ \citenamefont {Chandra}}]{Ghosh}%
  \BibitemOpen
  \bibfield  {author} {\bibinfo {author} {\bibfnamefont {K.}~\bibnamefont
  {Ghosh}}, \bibinfo {author} {\bibfnamefont {S.}~\bibnamefont
  {Ramakrishnan}},\ and\ \bibinfo {author} {\bibfnamefont {G.}~\bibnamefont
  {Chandra}},\ }\href {https://doi.org/10.1103/PhysRevB.48.4152} {\bibfield
  {journal} {\bibinfo  {journal} {Phys. Rev. B}\ }\textbf {\bibinfo {volume}
  {48}},\ \bibinfo {pages} {4152} (\bibinfo {year} {1993})}\BibitemShut
  {NoStop}%
\bibitem [{\citenamefont {Becker}\ \emph {et~al.}(1999)\citenamefont {Becker},
  \citenamefont {Patil}, \citenamefont {Ramakrishnan}, \citenamefont
  {Menovsky}, \citenamefont {Nieuwenhuys}, \citenamefont {Mydosh},
  \citenamefont {Kohgi},\ and\ \citenamefont {Iwasa}}]{Becker}%
  \BibitemOpen
  \bibfield  {author} {\bibinfo {author} {\bibfnamefont {B.}~\bibnamefont
  {Becker}}, \bibinfo {author} {\bibfnamefont {N.~G.}\ \bibnamefont {Patil}},
  \bibinfo {author} {\bibfnamefont {S.}~\bibnamefont {Ramakrishnan}}, \bibinfo
  {author} {\bibfnamefont {A.~A.}\ \bibnamefont {Menovsky}}, \bibinfo {author}
  {\bibfnamefont {G.~J.}\ \bibnamefont {Nieuwenhuys}}, \bibinfo {author}
  {\bibfnamefont {J.~A.}\ \bibnamefont {Mydosh}}, \bibinfo {author}
  {\bibfnamefont {M.}~\bibnamefont {Kohgi}},\ and\ \bibinfo {author}
  {\bibfnamefont {K.}~\bibnamefont {Iwasa}},\ }\href
  {https://doi.org/10.1103/PhysRevB.59.7266} {\bibfield  {journal} {\bibinfo
  {journal} {Phys. Rev. B}\ }\textbf {\bibinfo {volume} {59}},\ \bibinfo
  {pages} {7266} (\bibinfo {year} {1999})}\BibitemShut {NoStop}%
\bibitem [{\citenamefont {Galli}\ \emph {et~al.}(2000)\citenamefont {Galli},
  \citenamefont {Ramakrishnan}, \citenamefont {Taniguchi}, \citenamefont
  {Nieuwenhuys}, \citenamefont {Mydosh}, \citenamefont {Geupel}, \citenamefont
  {L{\"u}decke},\ and\ \citenamefont {van Smaalen}}]{Galli}%
  \BibitemOpen
  \bibfield  {author} {\bibinfo {author} {\bibfnamefont {F.}~\bibnamefont
  {Galli}}, \bibinfo {author} {\bibfnamefont {S.}~\bibnamefont {Ramakrishnan}},
  \bibinfo {author} {\bibfnamefont {T.}~\bibnamefont {Taniguchi}}, \bibinfo
  {author} {\bibfnamefont {G.~J.}\ \bibnamefont {Nieuwenhuys}}, \bibinfo
  {author} {\bibfnamefont {J.}~\bibnamefont {Mydosh}}, \bibinfo {author}
  {\bibfnamefont {S.}~\bibnamefont {Geupel}}, \bibinfo {author} {\bibfnamefont
  {J.}~\bibnamefont {L{\"u}decke}},\ and\ \bibinfo {author} {\bibfnamefont
  {S.}~\bibnamefont {van Smaalen}},\ }\href
  {https://doi.org/10.1103/PhysRevLett.85.158} {\bibfield  {journal} {\bibinfo
  {journal} {Phys. Rev. Lett.}\ }\textbf {\bibinfo {volume} {85}},\ \bibinfo
  {pages} {158} (\bibinfo {year} {2000})}\BibitemShut {NoStop}%
\bibitem [{\citenamefont {Kuo}\ \emph {et~al.}(2003)\citenamefont {Kuo},
  \citenamefont {Hsu}, \citenamefont {Li}, \citenamefont {Huang}, \citenamefont
  {Huang}, \citenamefont {Lue},\ and\ \citenamefont {Yang}}]{Kuo}%
  \BibitemOpen
  \bibfield  {author} {\bibinfo {author} {\bibfnamefont {Y.-K.}\ \bibnamefont
  {Kuo}}, \bibinfo {author} {\bibfnamefont {F.}~\bibnamefont {Hsu}}, \bibinfo
  {author} {\bibfnamefont {H.}~\bibnamefont {Li}}, \bibinfo {author}
  {\bibfnamefont {H.}~\bibnamefont {Huang}}, \bibinfo {author} {\bibfnamefont
  {C.}~\bibnamefont {Huang}}, \bibinfo {author} {\bibfnamefont
  {C.}~\bibnamefont {Lue}},\ and\ \bibinfo {author} {\bibfnamefont
  {H.}~\bibnamefont {Yang}},\ }\href
  {https://doi.org/10.1103/PhysRevB.67.195101} {\bibfield  {journal} {\bibinfo
  {journal} {Phys. Rev. B}\ }\textbf {\bibinfo {volume} {67}},\ \bibinfo
  {pages} {195101} (\bibinfo {year} {2003})}\BibitemShut {NoStop}%
\bibitem [{\citenamefont {van Smaalen}\ \emph {et~al.}(2004)\citenamefont {van
  Smaalen}, \citenamefont {Shaz}, \citenamefont {Palatinus}, \citenamefont
  {Daniels}, \citenamefont {Galli}, \citenamefont {Nieuwenhuys},\ and\
  \citenamefont {Mydosh}}]{Smaalen}%
  \BibitemOpen
  \bibfield  {author} {\bibinfo {author} {\bibfnamefont {S.}~\bibnamefont {van
  Smaalen}}, \bibinfo {author} {\bibfnamefont {M.}~\bibnamefont {Shaz}},
  \bibinfo {author} {\bibfnamefont {L.}~\bibnamefont {Palatinus}}, \bibinfo
  {author} {\bibfnamefont {P.}~\bibnamefont {Daniels}}, \bibinfo {author}
  {\bibfnamefont {F.}~\bibnamefont {Galli}}, \bibinfo {author} {\bibfnamefont
  {G.~J.}\ \bibnamefont {Nieuwenhuys}},\ and\ \bibinfo {author} {\bibfnamefont
  {J.~A.}\ \bibnamefont {Mydosh}},\ }\href
  {https://doi.org/10.1103/PhysRevB.69.014103} {\bibfield  {journal} {\bibinfo
  {journal} {Phys. Rev. B}\ }\textbf {\bibinfo {volume} {69}},\ \bibinfo
  {pages} {014103} (\bibinfo {year} {2004})}\BibitemShut {NoStop}%
\bibitem [{\citenamefont {Hausermann-Berg}\ and\ \citenamefont
  {Shelton}(1987)}]{Hausermann}%
  \BibitemOpen
  \bibfield  {author} {\bibinfo {author} {\bibfnamefont {L.~S.}\ \bibnamefont
  {Hausermann-Berg}}\ and\ \bibinfo {author} {\bibfnamefont {R.~N.}\
  \bibnamefont {Shelton}},\ }\href {https://doi.org/10.1103/PhysRevB.35.6659}
  {\bibfield  {journal} {\bibinfo  {journal} {Phys. Rev. B}\ }\textbf {\bibinfo
  {volume} {35}},\ \bibinfo {pages} {6659} (\bibinfo {year}
  {1987})}\BibitemShut {NoStop}%
\bibitem [{\citenamefont {Matthias}\ and\ \citenamefont
  {Hulm}(1953)}]{Matthias}%
  \BibitemOpen
  \bibfield  {author} {\bibinfo {author} {\bibfnamefont {B.~T.}\ \bibnamefont
  {Matthias}}\ and\ \bibinfo {author} {\bibfnamefont {J.~K.}\ \bibnamefont
  {Hulm}},\ }\href {https://doi.org/10.1103/PhysRev.89.439} {\bibfield
  {journal} {\bibinfo  {journal} {Phys. Rev.}\ }\textbf {\bibinfo {volume}
  {89}},\ \bibinfo {pages} {439} (\bibinfo {year} {1953})}\BibitemShut
  {NoStop}%
\bibitem [{\citenamefont {Rai}\ \emph {et~al.}(2015)\citenamefont {Rai},
  \citenamefont {Oswald}, \citenamefont {Wang}, \citenamefont {McCandless},
  \citenamefont {Chan},\ and\ \citenamefont {Morosan}}]{rai2015}%
  \BibitemOpen
  \bibfield  {author} {\bibinfo {author} {\bibfnamefont {B.~K.}\ \bibnamefont
  {Rai}}, \bibinfo {author} {\bibfnamefont {I.~W.~H.}\ \bibnamefont {Oswald}},
  \bibinfo {author} {\bibfnamefont {J.~K.}\ \bibnamefont {Wang}}, \bibinfo
  {author} {\bibfnamefont {G.~T.}\ \bibnamefont {McCandless}}, \bibinfo
  {author} {\bibfnamefont {J.~Y.}\ \bibnamefont {Chan}},\ and\ \bibinfo
  {author} {\bibfnamefont {E.}~\bibnamefont {Morosan}},\ }\href
  {https://doi.org/10.1021/cm504658h} {\bibfield  {journal} {\bibinfo
  {journal} {Chem. Mater.}\ }\textbf {\bibinfo {volume} {27}},\ \bibinfo
  {pages} {2488} (\bibinfo {year} {2015})}\BibitemShut {NoStop}%
\bibitem [{\citenamefont {Takada}\ \emph {et~al.}(2003)\citenamefont {Takada},
  \citenamefont {Sakurai}, \citenamefont {Takayama-Muromachi}, \citenamefont
  {Izumi}, \citenamefont {Dilanian},\ and\ \citenamefont
  {Sasaki}}]{Takada2003}%
  \BibitemOpen
  \bibfield  {author} {\bibinfo {author} {\bibfnamefont {K.}~\bibnamefont
  {Takada}}, \bibinfo {author} {\bibfnamefont {H.}~\bibnamefont {Sakurai}},
  \bibinfo {author} {\bibfnamefont {E.}~\bibnamefont {Takayama-Muromachi}},
  \bibinfo {author} {\bibfnamefont {F.}~\bibnamefont {Izumi}}, \bibinfo
  {author} {\bibfnamefont {R.~A.}\ \bibnamefont {Dilanian}},\ and\ \bibinfo
  {author} {\bibfnamefont {T.}~\bibnamefont {Sasaki}},\ }\href
  {https://doi.org/10.1038/nature01450} {\bibfield  {journal} {\bibinfo
  {journal} {Nature}\ }\textbf {\bibinfo {volume} {422}},\ \bibinfo {pages}
  {53} (\bibinfo {year} {2003})}\BibitemShut {NoStop}%
\bibitem [{\citenamefont {He}\ \emph {et~al.}(2021)\citenamefont {He},
  \citenamefont {Huang}, \citenamefont {Zhou}, \citenamefont {Liu},
  \citenamefont {Guo}, \citenamefont {Song}, \citenamefont {Guo}, \citenamefont
  {Hu}, \citenamefont {He}, \citenamefont {Huang}, \citenamefont {Li},
  \citenamefont {Zhang}, \citenamefont {Wang}, \citenamefont {Guo},
  \citenamefont {Xing},\ and\ \citenamefont {Chen}}]{He2021}%
  \BibitemOpen
  \bibfield  {author} {\bibinfo {author} {\bibfnamefont {Z.}~\bibnamefont
  {He}}, \bibinfo {author} {\bibfnamefont {R.}~\bibnamefont {Huang}}, \bibinfo
  {author} {\bibfnamefont {K.}~\bibnamefont {Zhou}}, \bibinfo {author}
  {\bibfnamefont {Y.}~\bibnamefont {Liu}}, \bibinfo {author} {\bibfnamefont
  {S.}~\bibnamefont {Guo}}, \bibinfo {author} {\bibfnamefont {Y.}~\bibnamefont
  {Song}}, \bibinfo {author} {\bibfnamefont {Z.}~\bibnamefont {Guo}}, \bibinfo
  {author} {\bibfnamefont {S.}~\bibnamefont {Hu}}, \bibinfo {author}
  {\bibfnamefont {L.}~\bibnamefont {He}}, \bibinfo {author} {\bibfnamefont
  {Q.}~\bibnamefont {Huang}}, \bibinfo {author} {\bibfnamefont
  {L.}~\bibnamefont {Li}}, \bibinfo {author} {\bibfnamefont {J.}~\bibnamefont
  {Zhang}}, \bibinfo {author} {\bibfnamefont {S.}~\bibnamefont {Wang}},
  \bibinfo {author} {\bibfnamefont {J.}~\bibnamefont {Guo}}, \bibinfo {author}
  {\bibfnamefont {X.}~\bibnamefont {Xing}},\ and\ \bibinfo {author}
  {\bibfnamefont {J.}~\bibnamefont {Chen}},\ }\href
  {https://doi.org/10.1021/acs.inorgchem.1c00699} {\bibfield  {journal}
  {\bibinfo  {journal} {Inorg. Chem.}\ }\textbf {\bibinfo {volume} {60}},\
  \bibinfo {pages} {6157} (\bibinfo {year} {2021})}\BibitemShut {NoStop}%
\bibitem [{\citenamefont {Scherer}\ \emph {et~al.}(2010)\citenamefont
  {Scherer}, \citenamefont {Hauf}, \citenamefont {Presnitz}, \citenamefont
  {Scheidt}, \citenamefont {Eickerling}, \citenamefont {Eyert}, \citenamefont
  {Hoffmann}, \citenamefont {Rodewald}, \citenamefont {Hammerschmidt},
  \citenamefont {Vogt},\ and\ \citenamefont {Pöttgen}}]{Scherer2010}%
  \BibitemOpen
  \bibfield  {author} {\bibinfo {author} {\bibfnamefont {W.}~\bibnamefont
  {Scherer}}, \bibinfo {author} {\bibfnamefont {C.}~\bibnamefont {Hauf}},
  \bibinfo {author} {\bibfnamefont {M.}~\bibnamefont {Presnitz}}, \bibinfo
  {author} {\bibfnamefont {E.-W.}\ \bibnamefont {Scheidt}}, \bibinfo {author}
  {\bibfnamefont {G.}~\bibnamefont {Eickerling}}, \bibinfo {author}
  {\bibfnamefont {V.}~\bibnamefont {Eyert}}, \bibinfo {author} {\bibfnamefont
  {R.-D.}\ \bibnamefont {Hoffmann}}, \bibinfo {author} {\bibfnamefont
  {U.}~\bibnamefont {Rodewald}}, \bibinfo {author} {\bibfnamefont
  {A.}~\bibnamefont {Hammerschmidt}}, \bibinfo {author} {\bibfnamefont
  {C.}~\bibnamefont {Vogt}},\ and\ \bibinfo {author} {\bibfnamefont
  {R.}~\bibnamefont {Pöttgen}},\ }\href
  {https://doi.org/https://doi.org/10.1002/anie.200904956} {\bibfield
  {journal} {\bibinfo  {journal} {Angew. Chem. Int. Edit.}\ }\textbf {\bibinfo
  {volume} {49}},\ \bibinfo {pages} {1578} (\bibinfo {year}
  {2010})}\BibitemShut {NoStop}%
\bibitem [{\citenamefont {Braun}\ \emph {et~al.}(1980)\citenamefont {Braun},
  \citenamefont {Yvon},\ and\ \citenamefont {Braun}}]{Braun80}%
  \BibitemOpen
  \bibfield  {author} {\bibinfo {author} {\bibfnamefont {H.~F.}\ \bibnamefont
  {Braun}}, \bibinfo {author} {\bibfnamefont {K.}~\bibnamefont {Yvon}},\ and\
  \bibinfo {author} {\bibfnamefont {R.~M.}\ \bibnamefont {Braun}},\ }\href
  {https://doi.org/10.1107/S0567740880008825} {\bibfield  {journal} {\bibinfo
  {journal} {Acta. Crystall. B-Stru.}\ }\textbf {\bibinfo {volume} {36}},\
  \bibinfo {pages} {2397} (\bibinfo {year} {1980})}\BibitemShut {NoStop}%
\bibitem [{\citenamefont {Fischer}(1978)}]{Fischer}%
  \BibitemOpen
  \bibfield  {author} {\bibinfo {author} {\bibfnamefont {{\O}.}~\bibnamefont
  {Fischer}},\ }\href {https://doi.org/10.1007/BF00931416} {\bibfield
  {journal} {\bibinfo  {journal} {Appl. Phys.}\ }\textbf {\bibinfo {volume}
  {16}},\ \bibinfo {pages} {1} (\bibinfo {year} {1978})}\BibitemShut {NoStop}%
\bibitem [{\citenamefont {Vandenberg}\ and\ \citenamefont
  {Matthias}(1977)}]{Vandenberg}%
  \BibitemOpen
  \bibfield  {author} {\bibinfo {author} {\bibfnamefont {J.~M.}\ \bibnamefont
  {Vandenberg}}\ and\ \bibinfo {author} {\bibfnamefont {B.~T.}\ \bibnamefont
  {Matthias}},\ }\href {https://doi.org/10.1126/science.198.4313.194}
  {\bibfield  {journal} {\bibinfo  {journal} {Science}\ }\textbf {\bibinfo
  {volume} {198}},\ \bibinfo {pages} {194} (\bibinfo {year}
  {1977})}\BibitemShut {NoStop}%
\bibitem [{\citenamefont {Nakajima}\ \emph {et~al.}(2009)\citenamefont
  {Nakajima}, \citenamefont {Li},\ and\ \citenamefont
  {Tamegai}}]{Nakajima2009}%
  \BibitemOpen
  \bibfield  {author} {\bibinfo {author} {\bibfnamefont {Y.}~\bibnamefont
  {Nakajima}}, \bibinfo {author} {\bibfnamefont {G.~J.}\ \bibnamefont {Li}},\
  and\ \bibinfo {author} {\bibfnamefont {T.}~\bibnamefont {Tamegai}},\ }\href
  {https://doi.org/10.1088/1742-6596/150/5/052180} {\bibfield  {journal}
  {\bibinfo  {journal} {J. Phys. Conf. Ser.}\ }\textbf {\bibinfo {volume}
  {150}},\ \bibinfo {pages} {052180} (\bibinfo {year} {2009})}\BibitemShut
  {NoStop}%
\bibitem [{\citenamefont {Koyama}\ \emph {et~al.}(1999)\citenamefont {Koyama},
  \citenamefont {Sugita}, \citenamefont {Wada},\ and\ \citenamefont
  {Tsutsumi}}]{Koyama}%
  \BibitemOpen
  \bibfield  {author} {\bibinfo {author} {\bibfnamefont {T.}~\bibnamefont
  {Koyama}}, \bibinfo {author} {\bibfnamefont {H.}~\bibnamefont {Sugita}},
  \bibinfo {author} {\bibfnamefont {S.}~\bibnamefont {Wada}},\ and\ \bibinfo
  {author} {\bibfnamefont {K.}~\bibnamefont {Tsutsumi}},\ }\href
  {https://doi.org/10.1143/JPSJ.68.2326} {\bibfield  {journal} {\bibinfo
  {journal} {J. Phys. Soc. Jpn.}\ }\textbf {\bibinfo {volume} {68}},\ \bibinfo
  {pages} {2326} (\bibinfo {year} {1999})}\BibitemShut {NoStop}%
\bibitem [{\citenamefont {Lue}\ \emph {et~al.}(2008)\citenamefont {Lue},
  \citenamefont {Liu}, \citenamefont {Fu}, \citenamefont {Cheng},\ and\
  \citenamefont {Yang}}]{Lue}%
  \BibitemOpen
  \bibfield  {author} {\bibinfo {author} {\bibfnamefont {C.~S.}\ \bibnamefont
  {Lue}}, \bibinfo {author} {\bibfnamefont {R.~F.}\ \bibnamefont {Liu}},
  \bibinfo {author} {\bibfnamefont {Y.~F.}\ \bibnamefont {Fu}}, \bibinfo
  {author} {\bibfnamefont {C.}~\bibnamefont {Cheng}},\ and\ \bibinfo {author}
  {\bibfnamefont {H.~D.}\ \bibnamefont {Yang}},\ }\href
  {https://doi.org/10.1103/PhysRevB.77.115130} {\bibfield  {journal} {\bibinfo
  {journal} {Phys. Rev. B}\ }\textbf {\bibinfo {volume} {77}},\ \bibinfo
  {pages} {115130} (\bibinfo {year} {2008})}\BibitemShut {NoStop}%
\bibitem [{\citenamefont {Lee}\ \emph {et~al.}(1999)\citenamefont {Lee},
  \citenamefont {Cywinski},\ and\ \citenamefont {Kilcoyne}}]{Lee1999}%
  \BibitemOpen
  \bibfield  {author} {\bibinfo {author} {\bibfnamefont {S.~L.}\ \bibnamefont
  {Lee}}, \bibinfo {author} {\bibfnamefont {R.}~\bibnamefont {Cywinski}},\ and\
  \bibinfo {author} {\bibfnamefont {S.~H.}\ \bibnamefont {Kilcoyne}},\
  }\href@noop {} {\emph {\bibinfo {title} {Muon Science: Muons in Physics,
  Chemistry and Materials}}},\ Vol.~\bibinfo {volume} {51}\ (\bibinfo
  {publisher} {SUSSP Publications and IOP Publishing},\ \bibinfo {address}
  {Bristol},\ \bibinfo {year} {1999})\BibitemShut {NoStop}%
\bibitem [{\citenamefont {Adroja}\ \emph {et~al.}(2018)\citenamefont {Adroja},
  \citenamefont {Biswas}, \citenamefont {Bhattacharyya}, \citenamefont
  {Hillier},\ and\ \citenamefont {Anand}}]{Bhattacharyya1321}%
  \BibitemOpen
  \bibfield  {author} {\bibinfo {author} {\bibfnamefont {D.~T.}\ \bibnamefont
  {Adroja}}, \bibinfo {author} {\bibfnamefont {P.~K.}\ \bibnamefont {Biswas}},
  \bibinfo {author} {\bibfnamefont {A.}~\bibnamefont {Bhattacharyya}}, \bibinfo
  {author} {\bibfnamefont {A.}~\bibnamefont {Hillier}},\ and\ \bibinfo {author}
  {\bibfnamefont {V.~K.}\ \bibnamefont {Anand}},\ }\bibfield  {journal}
  {\bibinfo  {journal} {https://doi.org/10.5286/ISIS.E.92918257}\ }\href
  {https://doi.org/10.5286/ISIS.E.RB1810871} {10.5286/ISIS.E.RB1810871}
  (\bibinfo {year} {2018})\BibitemShut {NoStop}%
\bibitem [{\citenamefont {Pratt}(2000)}]{Pratt2000}%
  \BibitemOpen
  \bibfield  {author} {\bibinfo {author} {\bibfnamefont {F.~L.}\ \bibnamefont
  {Pratt}},\ }\href {https://doi.org/10.1016/S0921-4526(00)00328-8} {\bibfield
  {journal} {\bibinfo  {journal} {Physica B}\ }\textbf {\bibinfo {volume}
  {289--290}},\ \bibinfo {pages} {710} (\bibinfo {year} {2000})}\BibitemShut
  {NoStop}%
\bibitem [{\citenamefont {Kresse}\ and\ \citenamefont
  {Hafner}(1993)}]{kresse1993}%
  \BibitemOpen
  \bibfield  {author} {\bibinfo {author} {\bibfnamefont {G.}~\bibnamefont
  {Kresse}}\ and\ \bibinfo {author} {\bibfnamefont {J.}~\bibnamefont
  {Hafner}},\ }\href {https://doi.org/10.1103/PhysRevB.47.558} {\bibfield
  {journal} {\bibinfo  {journal} {Phys. Rev. B}\ }\textbf {\bibinfo {volume}
  {47}},\ \bibinfo {pages} {558} (\bibinfo {year} {1993})}\BibitemShut
  {NoStop}%
\bibitem [{\citenamefont {Kresse}\ and\ \citenamefont
  {Furthm\"uller}(1996)}]{kresse1996}%
  \BibitemOpen
  \bibfield  {author} {\bibinfo {author} {\bibfnamefont {G.}~\bibnamefont
  {Kresse}}\ and\ \bibinfo {author} {\bibfnamefont {J.}~\bibnamefont
  {Furthm\"uller}},\ }\href {https://doi.org/10.1103/PhysRevB.54.11169}
  {\bibfield  {journal} {\bibinfo  {journal} {Phys. Rev. B}\ }\textbf {\bibinfo
  {volume} {54}},\ \bibinfo {pages} {11169} (\bibinfo {year}
  {1996})}\BibitemShut {NoStop}%
\bibitem [{\citenamefont {Kresse}\ and\ \citenamefont
  {Furthm{\"u}ller}(1996)}]{kresse1996_2}%
  \BibitemOpen
  \bibfield  {author} {\bibinfo {author} {\bibfnamefont {G.}~\bibnamefont
  {Kresse}}\ and\ \bibinfo {author} {\bibfnamefont {J.}~\bibnamefont
  {Furthm{\"u}ller}},\ }\href
  {https://doi.org/https://doi.org/10.1016/0927-0256(96)00008-0} {\bibfield
  {journal} {\bibinfo  {journal} {Comp. Mater. Sci.}\ }\textbf {\bibinfo
  {volume} {6}},\ \bibinfo {pages} {15} (\bibinfo {year} {1996})}\BibitemShut
  {NoStop}%
\bibitem [{\citenamefont {Bl\"ochl}(1994)}]{blochl1994}%
  \BibitemOpen
  \bibfield  {author} {\bibinfo {author} {\bibfnamefont {P.~E.}\ \bibnamefont
  {Bl\"ochl}},\ }\href {https://doi.org/10.1103/PhysRevB.50.17953} {\bibfield
  {journal} {\bibinfo  {journal} {Phys. Rev. B}\ }\textbf {\bibinfo {volume}
  {50}},\ \bibinfo {pages} {17953} (\bibinfo {year} {1994})}\BibitemShut
  {NoStop}%
\bibitem [{\citenamefont {Kresse}\ and\ \citenamefont
  {Joubert}(1999)}]{kresse1999}%
  \BibitemOpen
  \bibfield  {author} {\bibinfo {author} {\bibfnamefont {G.}~\bibnamefont
  {Kresse}}\ and\ \bibinfo {author} {\bibfnamefont {D.}~\bibnamefont
  {Joubert}},\ }\href {https://doi.org/10.1103/PhysRevB.59.1758} {\bibfield
  {journal} {\bibinfo  {journal} {Phys. Rev. B}\ }\textbf {\bibinfo {volume}
  {59}},\ \bibinfo {pages} {1758} (\bibinfo {year} {1999})}\BibitemShut
  {NoStop}%
\bibitem [{\citenamefont {Perdew}\ \emph {et~al.}(1996)\citenamefont {Perdew},
  \citenamefont {Burke},\ and\ \citenamefont {Ernzerhof}}]{perdew1996}%
  \BibitemOpen
  \bibfield  {author} {\bibinfo {author} {\bibfnamefont {J.~P.}\ \bibnamefont
  {Perdew}}, \bibinfo {author} {\bibfnamefont {K.}~\bibnamefont {Burke}},\ and\
  \bibinfo {author} {\bibfnamefont {M.}~\bibnamefont {Ernzerhof}},\ }\href
  {https://doi.org/10.1103/PhysRevLett.77.3865} {\bibfield  {journal} {\bibinfo
   {journal} {Phys. Rev. Lett.}\ }\textbf {\bibinfo {volume} {77}},\ \bibinfo
  {pages} {3865} (\bibinfo {year} {1996})}\BibitemShut {NoStop}%
\bibitem [{\citenamefont {Singh}\ \emph {et~al.}(2005)\citenamefont {Singh},
  \citenamefont {Nirmala}, \citenamefont {Ramakrishnan},\ and\ \citenamefont
  {Malik}}]{Singh}%
  \BibitemOpen
  \bibfield  {author} {\bibinfo {author} {\bibfnamefont {Y.}~\bibnamefont
  {Singh}}, \bibinfo {author} {\bibfnamefont {R.}~\bibnamefont {Nirmala}},
  \bibinfo {author} {\bibfnamefont {S.}~\bibnamefont {Ramakrishnan}},\ and\
  \bibinfo {author} {\bibfnamefont {S.}~\bibnamefont {Malik}},\ }\href@noop {}
  {\bibfield  {journal} {\bibinfo  {journal} {Physical Review B}\ }\textbf
  {\bibinfo {volume} {72}},\ \bibinfo {pages} {045106} (\bibinfo {year}
  {2005})}\BibitemShut {NoStop}%
\bibitem [{\citenamefont {Werthamer}\ \emph {et~al.}(1966)\citenamefont
  {Werthamer}, \citenamefont {Helfand},\ and\ \citenamefont {Hohenberg}}]{WHH}%
  \BibitemOpen
  \bibfield  {author} {\bibinfo {author} {\bibfnamefont {N.~R.}\ \bibnamefont
  {Werthamer}}, \bibinfo {author} {\bibfnamefont {E.}~\bibnamefont {Helfand}},\
  and\ \bibinfo {author} {\bibfnamefont {P.~C.}\ \bibnamefont {Hohenberg}},\
  }\href {https://doi.org/10.1103/PhysRev.147.295} {\bibfield  {journal}
  {\bibinfo  {journal} {Phys. Rev.}\ }\textbf {\bibinfo {volume} {147}},\
  \bibinfo {pages} {295} (\bibinfo {year} {1966})}\BibitemShut {NoStop}%
\bibitem [{\citenamefont {Bardeen}\ \emph {et~al.}(1957)\citenamefont
  {Bardeen}, \citenamefont {Cooper},\ and\ \citenamefont {Schrieffer}}]{BCS}%
  \BibitemOpen
  \bibfield  {author} {\bibinfo {author} {\bibfnamefont {J.}~\bibnamefont
  {Bardeen}}, \bibinfo {author} {\bibfnamefont {L.~N.}\ \bibnamefont
  {Cooper}},\ and\ \bibinfo {author} {\bibfnamefont {J.~R.}\ \bibnamefont
  {Schrieffer}},\ }\href {https://doi.org/10.1103/PhysRev.108.1175} {\bibfield
  {journal} {\bibinfo  {journal} {Phys. Rev.}\ }\textbf {\bibinfo {volume}
  {108}},\ \bibinfo {pages} {1175} (\bibinfo {year} {1957})}\BibitemShut
  {NoStop}%
\bibitem [{\citenamefont {Bhattacharyya}\ \emph
  {et~al.}(2018{\natexlab{a}})\citenamefont {Bhattacharyya}, \citenamefont
  {Adroja}, \citenamefont {Smidman},\ and\ \citenamefont
  {Anand}}]{Bhattacharyyarev}%
  \BibitemOpen
  \bibfield  {author} {\bibinfo {author} {\bibfnamefont {A.}~\bibnamefont
  {Bhattacharyya}}, \bibinfo {author} {\bibfnamefont {D.~T.}\ \bibnamefont
  {Adroja}}, \bibinfo {author} {\bibfnamefont {M.}~\bibnamefont {Smidman}},\
  and\ \bibinfo {author} {\bibfnamefont {V.~K.}\ \bibnamefont {Anand}},\ }\href
  {https://doi.org/10.1007/s11433-018-9292-0} {\bibfield  {journal} {\bibinfo
  {journal} {Sci. China Phys. Mech. Astron.}\ }\textbf {\bibinfo {volume}
  {61}},\ \bibinfo {pages} {1} (\bibinfo {year}
  {2018}{\natexlab{a}})}\BibitemShut {NoStop}%
\bibitem [{\citenamefont {Bhattacharyya}\ \emph
  {et~al.}(2019{\natexlab{a}})\citenamefont {Bhattacharyya}, \citenamefont
  {Adroja}, \citenamefont {Panda}, \citenamefont {Saha}, \citenamefont {Das},
  \citenamefont {Machado}, \citenamefont {Cigarroa}, \citenamefont {Grant},
  \citenamefont {Fisk}, \citenamefont {Hillier},\ and\ \citenamefont
  {Manfrinetti}}]{BhattacharyyaThCoC2}%
  \BibitemOpen
  \bibfield  {author} {\bibinfo {author} {\bibfnamefont {A.}~\bibnamefont
  {Bhattacharyya}}, \bibinfo {author} {\bibfnamefont {D.~T.}\ \bibnamefont
  {Adroja}}, \bibinfo {author} {\bibfnamefont {K.}~\bibnamefont {Panda}},
  \bibinfo {author} {\bibfnamefont {S.}~\bibnamefont {Saha}}, \bibinfo {author}
  {\bibfnamefont {T.}~\bibnamefont {Das}}, \bibinfo {author} {\bibfnamefont
  {A.~J.~S.}\ \bibnamefont {Machado}}, \bibinfo {author} {\bibfnamefont
  {O.~V.}\ \bibnamefont {Cigarroa}}, \bibinfo {author} {\bibfnamefont {T.~W.}\
  \bibnamefont {Grant}}, \bibinfo {author} {\bibfnamefont {Z.}~\bibnamefont
  {Fisk}}, \bibinfo {author} {\bibfnamefont {A.~D.}\ \bibnamefont {Hillier}},\
  and\ \bibinfo {author} {\bibfnamefont {P.}~\bibnamefont {Manfrinetti}},\
  }\href {https://doi.org/10.1103/PhysRevLett.122.147001} {\bibfield  {journal}
  {\bibinfo  {journal} {Phys. Rev. Lett.}\ }\textbf {\bibinfo {volume} {122}},\
  \bibinfo {pages} {147001} (\bibinfo {year} {2019}{\natexlab{a}})}\BibitemShut
  {NoStop}%
\bibitem [{\citenamefont {Adroja}\ \emph {et~al.}(2021)\citenamefont {Adroja},
  \citenamefont {Bhattacharyya}, \citenamefont {Sato}, \citenamefont {Lees},
  \citenamefont {Biswas}, \citenamefont {Panda}, \citenamefont {Anand},
  \citenamefont {Stenning}, \citenamefont {Hillier},\ and\ \citenamefont
  {Aoki}}]{CeIr3}%
  \BibitemOpen
  \bibfield  {author} {\bibinfo {author} {\bibfnamefont {D.~T.}\ \bibnamefont
  {Adroja}}, \bibinfo {author} {\bibfnamefont {A.}~\bibnamefont
  {Bhattacharyya}}, \bibinfo {author} {\bibfnamefont {Y.~J.}\ \bibnamefont
  {Sato}}, \bibinfo {author} {\bibfnamefont {M.~R.}\ \bibnamefont {Lees}},
  \bibinfo {author} {\bibfnamefont {P.~K.}\ \bibnamefont {Biswas}}, \bibinfo
  {author} {\bibfnamefont {K.}~\bibnamefont {Panda}}, \bibinfo {author}
  {\bibfnamefont {V.~K.}\ \bibnamefont {Anand}}, \bibinfo {author}
  {\bibfnamefont {G.~B.~G.}\ \bibnamefont {Stenning}}, \bibinfo {author}
  {\bibfnamefont {A.~D.}\ \bibnamefont {Hillier}},\ and\ \bibinfo {author}
  {\bibfnamefont {D.}~\bibnamefont {Aoki}},\ }\href
  {https://doi.org/10.1103/PhysRevB.103.104514} {\bibfield  {journal} {\bibinfo
   {journal} {Phys. Rev. B}\ }\textbf {\bibinfo {volume} {103}},\ \bibinfo
  {pages} {104514} (\bibinfo {year} {2021})}\BibitemShut {NoStop}%
\bibitem [{\citenamefont {Bhattacharyya}\ \emph {et~al.}(2021)\citenamefont
  {Bhattacharyya}, \citenamefont {Ferreira}, \citenamefont {Panda},
  \citenamefont {Masunaga}, \citenamefont {de~Faria}, \citenamefont {Correa},
  \citenamefont {Santos}, \citenamefont {Adroja}, \citenamefont {Yokoyama},
  \citenamefont {Dorini}, \citenamefont {Jardim}, \citenamefont {Eleno},\ and\
  \citenamefont {Machado}}]{Zr5Pt3}%
  \BibitemOpen
  \bibfield  {author} {\bibinfo {author} {\bibfnamefont {A.}~\bibnamefont
  {Bhattacharyya}}, \bibinfo {author} {\bibfnamefont {P.~P.}\ \bibnamefont
  {Ferreira}}, \bibinfo {author} {\bibfnamefont {K.}~\bibnamefont {Panda}},
  \bibinfo {author} {\bibfnamefont {S.~H.}\ \bibnamefont {Masunaga}}, \bibinfo
  {author} {\bibfnamefont {L.~R.}\ \bibnamefont {de~Faria}}, \bibinfo {author}
  {\bibfnamefont {L.~E.}\ \bibnamefont {Correa}}, \bibinfo {author}
  {\bibfnamefont {F.~B.}\ \bibnamefont {Santos}}, \bibinfo {author}
  {\bibfnamefont {D.~T.}\ \bibnamefont {Adroja}}, \bibinfo {author}
  {\bibfnamefont {K.}~\bibnamefont {Yokoyama}}, \bibinfo {author}
  {\bibfnamefont {T.~T.}\ \bibnamefont {Dorini}}, \bibinfo {author}
  {\bibfnamefont {R.~F.}\ \bibnamefont {Jardim}}, \bibinfo {author}
  {\bibfnamefont {L.~T.~F.}\ \bibnamefont {Eleno}},\ and\ \bibinfo {author}
  {\bibfnamefont {A.~J.~S.}\ \bibnamefont {Machado}},\ }\href
  {https://doi.org/10.1088/1361-648x/ac2bc7} {\bibfield  {journal} {\bibinfo
  {journal} {J. Phys.: Condens. Matter}\ }\textbf {\bibinfo {volume} {34}},\
  \bibinfo {pages} {035602} (\bibinfo {year} {2021})}\BibitemShut {NoStop}%
\bibitem [{\citenamefont {Prozorov}\ and\ \citenamefont
  {Giannetta}(2006)}]{Prozorov}%
  \BibitemOpen
  \bibfield  {author} {\bibinfo {author} {\bibfnamefont {R.}~\bibnamefont
  {Prozorov}}\ and\ \bibinfo {author} {\bibfnamefont {R.~W.}\ \bibnamefont
  {Giannetta}},\ }\href {https://doi.org/10.1088/0953-2048/19/8/r01} {\bibfield
   {journal} {\bibinfo  {journal} {Supercond. Sci. Tech.}\ }\textbf {\bibinfo
  {volume} {19}},\ \bibinfo {pages} {R41} (\bibinfo {year} {2006})}\BibitemShut
  {NoStop}%
\bibitem [{\citenamefont {Adroja}\ \emph {et~al.}(2015)\citenamefont {Adroja},
  \citenamefont {Bhattacharyya}, \citenamefont {Telling}, \citenamefont {Feng},
  \citenamefont {Smidman}, \citenamefont {Pan}, \citenamefont {Zhao},
  \citenamefont {Hillier}, \citenamefont {Pratt},\ and\ \citenamefont
  {Strydom}}]{AdrojaK2Cr3As3}%
  \BibitemOpen
  \bibfield  {author} {\bibinfo {author} {\bibfnamefont {D.~T.}\ \bibnamefont
  {Adroja}}, \bibinfo {author} {\bibfnamefont {A.}~\bibnamefont
  {Bhattacharyya}}, \bibinfo {author} {\bibfnamefont {M.}~\bibnamefont
  {Telling}}, \bibinfo {author} {\bibfnamefont {Y.}~\bibnamefont {Feng}},
  \bibinfo {author} {\bibfnamefont {M.}~\bibnamefont {Smidman}}, \bibinfo
  {author} {\bibfnamefont {B.}~\bibnamefont {Pan}}, \bibinfo {author}
  {\bibfnamefont {J.}~\bibnamefont {Zhao}}, \bibinfo {author} {\bibfnamefont
  {A.~D.}\ \bibnamefont {Hillier}}, \bibinfo {author} {\bibfnamefont {F.~L.}\
  \bibnamefont {Pratt}},\ and\ \bibinfo {author} {\bibfnamefont {A.~M.}\
  \bibnamefont {Strydom}},\ }\href {https://doi.org/10.1103/PhysRevB.92.134505}
  {\bibfield  {journal} {\bibinfo  {journal} {Phys. Rev. B}\ }\textbf {\bibinfo
  {volume} {92}},\ \bibinfo {pages} {134505} (\bibinfo {year}
  {2015})}\BibitemShut {NoStop}%
\bibitem [{\citenamefont {Panda}\ \emph {et~al.}(2019)\citenamefont {Panda},
  \citenamefont {Bhattacharyya}, \citenamefont {Adroja}, \citenamefont {Kase},
  \citenamefont {Biswas}, \citenamefont {Saha}, \citenamefont {Das},
  \citenamefont {Lees},\ and\ \citenamefont {Hillier}}]{ZrIrSi}%
  \BibitemOpen
  \bibfield  {author} {\bibinfo {author} {\bibfnamefont {K.}~\bibnamefont
  {Panda}}, \bibinfo {author} {\bibfnamefont {A.}~\bibnamefont
  {Bhattacharyya}}, \bibinfo {author} {\bibfnamefont {D.~T.}\ \bibnamefont
  {Adroja}}, \bibinfo {author} {\bibfnamefont {N.}~\bibnamefont {Kase}},
  \bibinfo {author} {\bibfnamefont {P.~K.}\ \bibnamefont {Biswas}}, \bibinfo
  {author} {\bibfnamefont {S.}~\bibnamefont {Saha}}, \bibinfo {author}
  {\bibfnamefont {T.}~\bibnamefont {Das}}, \bibinfo {author} {\bibfnamefont
  {M.~R.}\ \bibnamefont {Lees}},\ and\ \bibinfo {author} {\bibfnamefont
  {A.~D.}\ \bibnamefont {Hillier}},\ }\href
  {https://doi.org/10.1103/PhysRevB.99.174513} {\bibfield  {journal} {\bibinfo
  {journal} {Phys. Rev. B}\ }\textbf {\bibinfo {volume} {99}},\ \bibinfo
  {pages} {174513} (\bibinfo {year} {2019})}\BibitemShut {NoStop}%
\bibitem [{\citenamefont {Pang}\ \emph {et~al.}(2015)\citenamefont {Pang},
  \citenamefont {Smidman}, \citenamefont {Jiang}, \citenamefont {Bao},
  \citenamefont {Weng}, \citenamefont {Wang}, \citenamefont {Jiao},
  \citenamefont {Zhang}, \citenamefont {Cao},\ and\ \citenamefont
  {Yuan}}]{Pang2015}%
  \BibitemOpen
  \bibfield  {author} {\bibinfo {author} {\bibfnamefont {G.~M.}\ \bibnamefont
  {Pang}}, \bibinfo {author} {\bibfnamefont {M.}~\bibnamefont {Smidman}},
  \bibinfo {author} {\bibfnamefont {W.~B.}\ \bibnamefont {Jiang}}, \bibinfo
  {author} {\bibfnamefont {J.~K.}\ \bibnamefont {Bao}}, \bibinfo {author}
  {\bibfnamefont {Z.~F.}\ \bibnamefont {Weng}}, \bibinfo {author}
  {\bibfnamefont {Y.~F.}\ \bibnamefont {Wang}}, \bibinfo {author}
  {\bibfnamefont {L.}~\bibnamefont {Jiao}}, \bibinfo {author} {\bibfnamefont
  {J.~L.}\ \bibnamefont {Zhang}}, \bibinfo {author} {\bibfnamefont {G.~H.}\
  \bibnamefont {Cao}},\ and\ \bibinfo {author} {\bibfnamefont {H.~Q.}\
  \bibnamefont {Yuan}},\ }\href {https://doi.org/10.1103/PhysRevB.91.220502}
  {\bibfield  {journal} {\bibinfo  {journal} {Phys. Rev. B}\ }\textbf {\bibinfo
  {volume} {91}},\ \bibinfo {pages} {220502} (\bibinfo {year}
  {2015})}\BibitemShut {NoStop}%
\bibitem [{\citenamefont {Annett}(1990)}]{Annet1990}%
  \BibitemOpen
  \bibfield  {author} {\bibinfo {author} {\bibfnamefont {J.~F.}\ \bibnamefont
  {Annett}},\ }\href {https://doi.org/10.1080/00018739000101481} {\bibfield
  {journal} {\bibinfo  {journal} {Adv. Phys.}\ }\textbf {\bibinfo {volume}
  {39}},\ \bibinfo {pages} {83} (\bibinfo {year} {1990})}\BibitemShut {NoStop}%
\bibitem [{\citenamefont {Brandt}(2003)}]{Brandt}%
  \BibitemOpen
  \bibfield  {author} {\bibinfo {author} {\bibfnamefont {E.~H.}\ \bibnamefont
  {Brandt}},\ }\href@noop {} {\bibfield  {journal} {\bibinfo  {journal}
  {Physical Review B}\ }\textbf {\bibinfo {volume} {68}},\ \bibinfo {pages}
  {054506} (\bibinfo {year} {2003})}\BibitemShut {NoStop}%
\bibitem [{\citenamefont {Chia}\ \emph {et~al.}(2004)\citenamefont {Chia},
  \citenamefont {Salamon}, \citenamefont {Sugawara},\ and\ \citenamefont
  {Sato}}]{Chia}%
  \BibitemOpen
  \bibfield  {author} {\bibinfo {author} {\bibfnamefont {E.~E.~M.}\
  \bibnamefont {Chia}}, \bibinfo {author} {\bibfnamefont {M.~B.}\ \bibnamefont
  {Salamon}}, \bibinfo {author} {\bibfnamefont {H.}~\bibnamefont {Sugawara}},\
  and\ \bibinfo {author} {\bibfnamefont {H.}~\bibnamefont {Sato}},\ }\href
  {https://doi.org/10.1103/PhysRevB.69.180509} {\bibfield  {journal} {\bibinfo
  {journal} {Phys. Rev. B}\ }\textbf {\bibinfo {volume} {69}},\ \bibinfo
  {pages} {180509} (\bibinfo {year} {2004})}\BibitemShut {NoStop}%
\bibitem [{\citenamefont {Amato}(1997)}]{Amato}%
  \BibitemOpen
  \bibfield  {author} {\bibinfo {author} {\bibfnamefont {A.}~\bibnamefont
  {Amato}},\ }\href {https://doi.org/10.1103/RevModPhys.69.1119} {\bibfield
  {journal} {\bibinfo  {journal} {Rev. Mod. Phys.}\ }\textbf {\bibinfo {volume}
  {69}},\ \bibinfo {pages} {1119} (\bibinfo {year} {1997})}\BibitemShut
  {NoStop}%
\bibitem [{\citenamefont {McMillan}(1968)}]{McMillan}%
  \BibitemOpen
  \bibfield  {author} {\bibinfo {author} {\bibfnamefont {W.~L.}\ \bibnamefont
  {McMillan}},\ }\href {https://doi.org/10.1103/PhysRev.167.331} {\bibfield
  {journal} {\bibinfo  {journal} {Phys. Rev.}\ }\textbf {\bibinfo {volume}
  {167}},\ \bibinfo {pages} {331} (\bibinfo {year} {1968})}\BibitemShut
  {NoStop}%
\bibitem [{\citenamefont {Bhattacharyya}\ \emph
  {et~al.}(2019{\natexlab{b}})\citenamefont {Bhattacharyya}, \citenamefont
  {Panda}, \citenamefont {Adroja}, \citenamefont {Kase}, \citenamefont
  {Biswas}, \citenamefont {Saha}, \citenamefont {Das}, \citenamefont {Lees},\
  and\ \citenamefont {Hillier}}]{HfIrSi}%
  \BibitemOpen
  \bibfield  {author} {\bibinfo {author} {\bibfnamefont {A.}~\bibnamefont
  {Bhattacharyya}}, \bibinfo {author} {\bibfnamefont {K.}~\bibnamefont
  {Panda}}, \bibinfo {author} {\bibfnamefont {D.~T.}\ \bibnamefont {Adroja}},
  \bibinfo {author} {\bibfnamefont {N.}~\bibnamefont {Kase}}, \bibinfo {author}
  {\bibfnamefont {P.~K.}\ \bibnamefont {Biswas}}, \bibinfo {author}
  {\bibfnamefont {S.}~\bibnamefont {Saha}}, \bibinfo {author} {\bibfnamefont
  {T.}~\bibnamefont {Das}}, \bibinfo {author} {\bibfnamefont {M.~R.}\
  \bibnamefont {Lees}},\ and\ \bibinfo {author} {\bibfnamefont {A.~D.}\
  \bibnamefont {Hillier}},\ }\href {https://doi.org/10.1088/1361-648x/ab549e}
  {\bibfield  {journal} {\bibinfo  {journal} {J. Phys.: Condens. Matter}\
  }\textbf {\bibinfo {volume} {32}},\ \bibinfo {pages} {085601} (\bibinfo
  {year} {2019}{\natexlab{b}})}\BibitemShut {NoStop}%
\bibitem [{\citenamefont {Bhattacharyya}\ \emph
  {et~al.}(2018{\natexlab{b}})\citenamefont {Bhattacharyya}, \citenamefont
  {Adroja}, \citenamefont {Kase}, \citenamefont {Hillier}, \citenamefont
  {Strydom},\ and\ \citenamefont {Akimitsu}}]{Sc5Rh6Sn18}%
  \BibitemOpen
  \bibfield  {author} {\bibinfo {author} {\bibfnamefont {A.}~\bibnamefont
  {Bhattacharyya}}, \bibinfo {author} {\bibfnamefont {D.~T.}\ \bibnamefont
  {Adroja}}, \bibinfo {author} {\bibfnamefont {N.}~\bibnamefont {Kase}},
  \bibinfo {author} {\bibfnamefont {A.~D.}\ \bibnamefont {Hillier}}, \bibinfo
  {author} {\bibfnamefont {A.~M.}\ \bibnamefont {Strydom}},\ and\ \bibinfo
  {author} {\bibfnamefont {J.}~\bibnamefont {Akimitsu}},\ }\href
  {https://doi.org/10.1103/PhysRevB.98.024511} {\bibfield  {journal} {\bibinfo
  {journal} {Phys. Rev. B}\ }\textbf {\bibinfo {volume} {98}},\ \bibinfo
  {pages} {024511} (\bibinfo {year} {2018}{\natexlab{b}})}\BibitemShut
  {NoStop}%
\bibitem [{\citenamefont {Luke}\ \emph {et~al.}(1998)\citenamefont {Luke},
  \citenamefont {Fudamoto}, \citenamefont {Kojima}, \citenamefont {Larkin},
  \citenamefont {Merrin}, \citenamefont {Nachumi}, \citenamefont {Uemura},
  \citenamefont {Maeno}, \citenamefont {Mao}, \citenamefont {Mori},
  \citenamefont {Nakamura},\ and\ \citenamefont {Sigrist}}]{Luke}%
  \BibitemOpen
  \bibfield  {author} {\bibinfo {author} {\bibfnamefont {G.~M.}\ \bibnamefont
  {Luke}}, \bibinfo {author} {\bibfnamefont {Y.}~\bibnamefont {Fudamoto}},
  \bibinfo {author} {\bibfnamefont {K.~M.}\ \bibnamefont {Kojima}}, \bibinfo
  {author} {\bibfnamefont {M.~I.}\ \bibnamefont {Larkin}}, \bibinfo {author}
  {\bibfnamefont {J.}~\bibnamefont {Merrin}}, \bibinfo {author} {\bibfnamefont
  {B.}~\bibnamefont {Nachumi}}, \bibinfo {author} {\bibfnamefont {Y.~J.}\
  \bibnamefont {Uemura}}, \bibinfo {author} {\bibfnamefont {Y.}~\bibnamefont
  {Maeno}}, \bibinfo {author} {\bibfnamefont {Z.~Q.}\ \bibnamefont {Mao}},
  \bibinfo {author} {\bibfnamefont {Y.}~\bibnamefont {Mori}}, \bibinfo {author}
  {\bibfnamefont {H.}~\bibnamefont {Nakamura}},\ and\ \bibinfo {author}
  {\bibfnamefont {M.}~\bibnamefont {Sigrist}},\ }\href
  {https://doi.org/10.1038/29038} {\bibfield  {journal} {\bibinfo  {journal}
  {Nature}\ }\textbf {\bibinfo {volume} {394}},\ \bibinfo {pages} {558}
  (\bibinfo {year} {1998})}\BibitemShut {NoStop}%
\bibitem [{\citenamefont {Luke}\ \emph {et~al.}(1993)\citenamefont {Luke},
  \citenamefont {Keren}, \citenamefont {Le}, \citenamefont {Wu}, \citenamefont
  {Uemura}, \citenamefont {Bonn}, \citenamefont {Taillefer},\ and\
  \citenamefont {Garrett}}]{luke1993muon}%
  \BibitemOpen
  \bibfield  {author} {\bibinfo {author} {\bibfnamefont {G.}~\bibnamefont
  {Luke}}, \bibinfo {author} {\bibfnamefont {A.}~\bibnamefont {Keren}},
  \bibinfo {author} {\bibfnamefont {L.}~\bibnamefont {Le}}, \bibinfo {author}
  {\bibfnamefont {W.}~\bibnamefont {Wu}}, \bibinfo {author} {\bibfnamefont
  {Y.}~\bibnamefont {Uemura}}, \bibinfo {author} {\bibfnamefont
  {D.}~\bibnamefont {Bonn}}, \bibinfo {author} {\bibfnamefont {L.}~\bibnamefont
  {Taillefer}},\ and\ \bibinfo {author} {\bibfnamefont {J.}~\bibnamefont
  {Garrett}},\ }\href@noop {} {\bibfield  {journal} {\bibinfo  {journal}
  {Physical review letters}\ }\textbf {\bibinfo {volume} {71}},\ \bibinfo
  {pages} {1466} (\bibinfo {year} {1993})}\BibitemShut {NoStop}%
\bibitem [{\citenamefont {Hillier}\ \emph {et~al.}(2009)\citenamefont
  {Hillier}, \citenamefont {Quintanilla},\ and\ \citenamefont
  {Cywinski}}]{LaNiC2}%
  \BibitemOpen
  \bibfield  {author} {\bibinfo {author} {\bibfnamefont {A.~D.}\ \bibnamefont
  {Hillier}}, \bibinfo {author} {\bibfnamefont {J.}~\bibnamefont
  {Quintanilla}},\ and\ \bibinfo {author} {\bibfnamefont {R.}~\bibnamefont
  {Cywinski}},\ }\href {https://doi.org/10.1103/PhysRevLett.102.117007}
  {\bibfield  {journal} {\bibinfo  {journal} {Phys. Rev. Lett.}\ }\textbf
  {\bibinfo {volume} {102}},\ \bibinfo {pages} {117007} (\bibinfo {year}
  {2009})}\BibitemShut {NoStop}%
\bibitem [{\citenamefont {Bhattacharyya}\ \emph
  {et~al.}(2015{\natexlab{a}})\citenamefont {Bhattacharyya}, \citenamefont
  {Adroja}, \citenamefont {Kase}, \citenamefont {Hillier}, \citenamefont
  {Akimitsu},\ and\ \citenamefont {Strydom}}]{Y5Rh6Sn18}%
  \BibitemOpen
  \bibfield  {author} {\bibinfo {author} {\bibfnamefont {A.}~\bibnamefont
  {Bhattacharyya}}, \bibinfo {author} {\bibfnamefont {D.}~\bibnamefont
  {Adroja}}, \bibinfo {author} {\bibfnamefont {N.}~\bibnamefont {Kase}},
  \bibinfo {author} {\bibfnamefont {A.}~\bibnamefont {Hillier}}, \bibinfo
  {author} {\bibfnamefont {J.}~\bibnamefont {Akimitsu}},\ and\ \bibinfo
  {author} {\bibfnamefont {A.}~\bibnamefont {Strydom}},\ }\href
  {https://doi.org/10.1038/srep12926} {\bibfield  {journal} {\bibinfo
  {journal} {Sci. Rep.}\ }\textbf {\bibinfo {volume} {5}},\ \bibinfo {pages}
  {12926} (\bibinfo {year} {2015}{\natexlab{a}})}\BibitemShut {NoStop}%
\bibitem [{\citenamefont {Bhattacharyya}\ \emph
  {et~al.}(2015{\natexlab{b}})\citenamefont {Bhattacharyya}, \citenamefont
  {Adroja}, \citenamefont {Quintanilla}, \citenamefont {Hillier}, \citenamefont
  {Kase}, \citenamefont {Strydom},\ and\ \citenamefont
  {Akimitsu}}]{Lu5Rh6Sn18}%
  \BibitemOpen
  \bibfield  {author} {\bibinfo {author} {\bibfnamefont {A.}~\bibnamefont
  {Bhattacharyya}}, \bibinfo {author} {\bibfnamefont {D.~T.}\ \bibnamefont
  {Adroja}}, \bibinfo {author} {\bibfnamefont {J.}~\bibnamefont {Quintanilla}},
  \bibinfo {author} {\bibfnamefont {A.~D.}\ \bibnamefont {Hillier}}, \bibinfo
  {author} {\bibfnamefont {N.}~\bibnamefont {Kase}}, \bibinfo {author}
  {\bibfnamefont {A.~M.}\ \bibnamefont {Strydom}},\ and\ \bibinfo {author}
  {\bibfnamefont {J.}~\bibnamefont {Akimitsu}},\ }\href
  {https://doi.org/10.1103/PhysRevB.91.060503} {\bibfield  {journal} {\bibinfo
  {journal} {Phys. Rev. B}\ }\textbf {\bibinfo {volume} {91}},\ \bibinfo
  {pages} {060503} (\bibinfo {year} {2015}{\natexlab{b}})}\BibitemShut
  {NoStop}%
\bibitem [{\citenamefont {Biswas}\ \emph {et~al.}(2013)\citenamefont {Biswas},
  \citenamefont {Luetkens}, \citenamefont {Neupert}, \citenamefont {St\"urzer},
  \citenamefont {Baines}, \citenamefont {Pascua}, \citenamefont {Schnyder},
  \citenamefont {Fischer}, \citenamefont {Goryo}, \citenamefont {Lees},
  \citenamefont {Maeter}, \citenamefont {Br\"uckner}, \citenamefont {Klauss},
  \citenamefont {Nicklas}, \citenamefont {Baker}, \citenamefont {Hillier},
  \citenamefont {Sigrist}, \citenamefont {Amato},\ and\ \citenamefont
  {Johrendt}}]{SrPtAs}%
  \BibitemOpen
  \bibfield  {author} {\bibinfo {author} {\bibfnamefont {P.~K.}\ \bibnamefont
  {Biswas}}, \bibinfo {author} {\bibfnamefont {H.}~\bibnamefont {Luetkens}},
  \bibinfo {author} {\bibfnamefont {T.}~\bibnamefont {Neupert}}, \bibinfo
  {author} {\bibfnamefont {T.}~\bibnamefont {St\"urzer}}, \bibinfo {author}
  {\bibfnamefont {C.}~\bibnamefont {Baines}}, \bibinfo {author} {\bibfnamefont
  {G.}~\bibnamefont {Pascua}}, \bibinfo {author} {\bibfnamefont {A.~P.}\
  \bibnamefont {Schnyder}}, \bibinfo {author} {\bibfnamefont {M.~H.}\
  \bibnamefont {Fischer}}, \bibinfo {author} {\bibfnamefont {J.}~\bibnamefont
  {Goryo}}, \bibinfo {author} {\bibfnamefont {M.~R.}\ \bibnamefont {Lees}},
  \bibinfo {author} {\bibfnamefont {H.}~\bibnamefont {Maeter}}, \bibinfo
  {author} {\bibfnamefont {F.}~\bibnamefont {Br\"uckner}}, \bibinfo {author}
  {\bibfnamefont {H.-H.}\ \bibnamefont {Klauss}}, \bibinfo {author}
  {\bibfnamefont {M.}~\bibnamefont {Nicklas}}, \bibinfo {author} {\bibfnamefont
  {P.~J.}\ \bibnamefont {Baker}}, \bibinfo {author} {\bibfnamefont {A.~D.}\
  \bibnamefont {Hillier}}, \bibinfo {author} {\bibfnamefont {M.}~\bibnamefont
  {Sigrist}}, \bibinfo {author} {\bibfnamefont {A.}~\bibnamefont {Amato}},\
  and\ \bibinfo {author} {\bibfnamefont {D.}~\bibnamefont {Johrendt}},\ }\href
  {https://doi.org/10.1103/PhysRevB.87.180503} {\bibfield  {journal} {\bibinfo
  {journal} {Phys. Rev. B}\ }\textbf {\bibinfo {volume} {87}},\ \bibinfo
  {pages} {180503} (\bibinfo {year} {2013})}\BibitemShut {NoStop}%
\bibitem [{\citenamefont {Barker}\ \emph {et~al.}(2015)\citenamefont {Barker},
  \citenamefont {Singh}, \citenamefont {Thamizhavel}, \citenamefont {Hillier},
  \citenamefont {Lees}, \citenamefont {Balakrishnan}, \citenamefont {Paul},\
  and\ \citenamefont {Singh}}]{Barker2015La7Ir3}%
  \BibitemOpen
  \bibfield  {author} {\bibinfo {author} {\bibfnamefont {J.~A.~T.}\
  \bibnamefont {Barker}}, \bibinfo {author} {\bibfnamefont {D.}~\bibnamefont
  {Singh}}, \bibinfo {author} {\bibfnamefont {A.}~\bibnamefont {Thamizhavel}},
  \bibinfo {author} {\bibfnamefont {A.~D.}\ \bibnamefont {Hillier}}, \bibinfo
  {author} {\bibfnamefont {M.~R.}\ \bibnamefont {Lees}}, \bibinfo {author}
  {\bibfnamefont {G.}~\bibnamefont {Balakrishnan}}, \bibinfo {author}
  {\bibfnamefont {D.~M.}\ \bibnamefont {Paul}},\ and\ \bibinfo {author}
  {\bibfnamefont {R.~P.}\ \bibnamefont {Singh}},\ }\href
  {https://doi.org/10.1103/PhysRevLett.115.267001} {\bibfield  {journal}
  {\bibinfo  {journal} {Phys. Rev. Lett.}\ }\textbf {\bibinfo {volume} {115}},\
  \bibinfo {pages} {267001} (\bibinfo {year} {2015})}\BibitemShut {NoStop}%
\bibitem [{\citenamefont {Singh}\ \emph {et~al.}(2020)\citenamefont {Singh},
  \citenamefont {Scheurer}, \citenamefont {Hillier}, \citenamefont {Adroja},\
  and\ \citenamefont {Singh}}]{Singh2018La7Rh3}%
  \BibitemOpen
  \bibfield  {author} {\bibinfo {author} {\bibfnamefont {D.}~\bibnamefont
  {Singh}}, \bibinfo {author} {\bibfnamefont {M.~S.}\ \bibnamefont {Scheurer}},
  \bibinfo {author} {\bibfnamefont {A.~D.}\ \bibnamefont {Hillier}}, \bibinfo
  {author} {\bibfnamefont {D.~T.}\ \bibnamefont {Adroja}},\ and\ \bibinfo
  {author} {\bibfnamefont {R.~P.}\ \bibnamefont {Singh}},\ }\href
  {https://doi.org/10.1103/PhysRevB.102.134511} {\bibfield  {journal} {\bibinfo
   {journal} {Phys. Rev. B}\ }\textbf {\bibinfo {volume} {102}},\ \bibinfo
  {pages} {134511} (\bibinfo {year} {2020})}\BibitemShut {NoStop}%
\bibitem [{\citenamefont {Mayoh}\ \emph {et~al.}(2021)\citenamefont {Mayoh},
  \citenamefont {Hillier}, \citenamefont {Balakrishnan},\ and\ \citenamefont
  {Lees}}]{Mayoh2021La7Pd3}%
  \BibitemOpen
  \bibfield  {author} {\bibinfo {author} {\bibfnamefont {D.~A.}\ \bibnamefont
  {Mayoh}}, \bibinfo {author} {\bibfnamefont {A.~D.}\ \bibnamefont {Hillier}},
  \bibinfo {author} {\bibfnamefont {G.}~\bibnamefont {Balakrishnan}},\ and\
  \bibinfo {author} {\bibfnamefont {M.~R.}\ \bibnamefont {Lees}},\ }\href
  {https://doi.org/10.1103/PhysRevB.103.024507} {\bibfield  {journal} {\bibinfo
   {journal} {Phys. Rev. B}\ }\textbf {\bibinfo {volume} {103}},\ \bibinfo
  {pages} {024507} (\bibinfo {year} {2021})}\BibitemShut {NoStop}%
\bibitem [{\citenamefont {K.~P.}\ \emph {et~al.}(2019)\citenamefont {K.~P.},
  \citenamefont {Singh}, \citenamefont {Biswas}, \citenamefont {Stenning},
  \citenamefont {Hillier},\ and\ \citenamefont {Singh}}]{Sajilesh2019Zr3Ir}%
  \BibitemOpen
  \bibfield  {author} {\bibinfo {author} {\bibfnamefont {S.}~\bibnamefont
  {K.~P.}}, \bibinfo {author} {\bibfnamefont {D.}~\bibnamefont {Singh}},
  \bibinfo {author} {\bibfnamefont {P.~K.}\ \bibnamefont {Biswas}}, \bibinfo
  {author} {\bibfnamefont {G.~B.~G.}\ \bibnamefont {Stenning}}, \bibinfo
  {author} {\bibfnamefont {A.~D.}\ \bibnamefont {Hillier}},\ and\ \bibinfo
  {author} {\bibfnamefont {R.~P.}\ \bibnamefont {Singh}},\ }\href
  {https://doi.org/10.1103/PhysRevMaterials.3.104802} {\bibfield  {journal}
  {\bibinfo  {journal} {Phys. Rev. Materials}\ }\textbf {\bibinfo {volume}
  {3}},\ \bibinfo {pages} {104802} (\bibinfo {year} {2019})}\BibitemShut
  {NoStop}%
\bibitem [{\citenamefont {Shang}\ \emph {et~al.}(2020)\citenamefont {Shang},
  \citenamefont {Ghosh}, \citenamefont {Zhao}, \citenamefont {Chang},
  \citenamefont {Baines}, \citenamefont {Lee}, \citenamefont {Gawryluk},
  \citenamefont {Shi}, \citenamefont {Medarde}, \citenamefont {Quintanilla},\
  and\ \citenamefont {Shiroka}}]{Shang2020Zr3Ir}%
  \BibitemOpen
  \bibfield  {author} {\bibinfo {author} {\bibfnamefont {T.}~\bibnamefont
  {Shang}}, \bibinfo {author} {\bibfnamefont {S.~K.}\ \bibnamefont {Ghosh}},
  \bibinfo {author} {\bibfnamefont {J.~Z.}\ \bibnamefont {Zhao}}, \bibinfo
  {author} {\bibfnamefont {L.-J.}\ \bibnamefont {Chang}}, \bibinfo {author}
  {\bibfnamefont {C.}~\bibnamefont {Baines}}, \bibinfo {author} {\bibfnamefont
  {M.~K.}\ \bibnamefont {Lee}}, \bibinfo {author} {\bibfnamefont {D.~J.}\
  \bibnamefont {Gawryluk}}, \bibinfo {author} {\bibfnamefont {M.}~\bibnamefont
  {Shi}}, \bibinfo {author} {\bibfnamefont {M.}~\bibnamefont {Medarde}},
  \bibinfo {author} {\bibfnamefont {J.}~\bibnamefont {Quintanilla}},\ and\
  \bibinfo {author} {\bibfnamefont {T.}~\bibnamefont {Shiroka}},\ }\href
  {https://doi.org/10.1103/PhysRevB.102.020503} {\bibfield  {journal} {\bibinfo
   {journal} {Phys. Rev. B}\ }\textbf {\bibinfo {volume} {102}},\ \bibinfo
  {pages} {020503} (\bibinfo {year} {2020})}\BibitemShut {NoStop}%
\bibitem [{\citenamefont {Luke}\ \emph {et~al.}(2000)\citenamefont {Luke},
  \citenamefont {Fudamoto}, \citenamefont {Kojima}, \citenamefont {Larkin},
  \citenamefont {Nachumi}, \citenamefont {Uemura}, \citenamefont {Sonier},
  \citenamefont {Maeno}, \citenamefont {Mao}, \citenamefont {Mori},\ and\
  \citenamefont {Agterberg}}]{Luke2000Sr2RuO4}%
  \BibitemOpen
  \bibfield  {author} {\bibinfo {author} {\bibfnamefont {G.~M.}\ \bibnamefont
  {Luke}}, \bibinfo {author} {\bibfnamefont {Y.}~\bibnamefont {Fudamoto}},
  \bibinfo {author} {\bibfnamefont {K.~M.}\ \bibnamefont {Kojima}}, \bibinfo
  {author} {\bibfnamefont {M.~I.}\ \bibnamefont {Larkin}}, \bibinfo {author}
  {\bibfnamefont {B.}~\bibnamefont {Nachumi}}, \bibinfo {author} {\bibfnamefont
  {Y.~J.}\ \bibnamefont {Uemura}}, \bibinfo {author} {\bibfnamefont {J.~E.}\
  \bibnamefont {Sonier}}, \bibinfo {author} {\bibfnamefont {Y.}~\bibnamefont
  {Maeno}}, \bibinfo {author} {\bibfnamefont {Z.~Q.}\ \bibnamefont {Mao}},
  \bibinfo {author} {\bibfnamefont {Y.}~\bibnamefont {Mori}},\ and\ \bibinfo
  {author} {\bibfnamefont {D.~F.}\ \bibnamefont {Agterberg}},\ }\href
  {https://doi.org/10.1016/S0921-4526(00)00414-2} {\bibfield  {journal}
  {\bibinfo  {journal} {Physica B}\ }\textbf {\bibinfo {volume} {289--290}},\
  \bibinfo {pages} {373} (\bibinfo {year} {2000})}\BibitemShut {NoStop}%
\bibitem [{\citenamefont {Xia}\ \emph {et~al.}(2006)\citenamefont {Xia},
  \citenamefont {Maeno}, \citenamefont {Beyersdorf}, \citenamefont {Fejer},\
  and\ \citenamefont {Kapitulnik}}]{Xia2006Sr2RuO4}%
  \BibitemOpen
  \bibfield  {author} {\bibinfo {author} {\bibfnamefont {J.}~\bibnamefont
  {Xia}}, \bibinfo {author} {\bibfnamefont {Y.}~\bibnamefont {Maeno}}, \bibinfo
  {author} {\bibfnamefont {P.~T.}\ \bibnamefont {Beyersdorf}}, \bibinfo
  {author} {\bibfnamefont {M.~M.}\ \bibnamefont {Fejer}},\ and\ \bibinfo
  {author} {\bibfnamefont {A.}~\bibnamefont {Kapitulnik}},\ }\href
  {https://doi.org/10.1103/PhysRevLett.97.167002} {\bibfield  {journal}
  {\bibinfo  {journal} {Phys. Rev. Lett.}\ }\textbf {\bibinfo {volume} {97}},\
  \bibinfo {pages} {167002} (\bibinfo {year} {2006})}\BibitemShut {NoStop}%
\bibitem [{\citenamefont {Kashiwaya}\ \emph {et~al.}(2019)\citenamefont
  {Kashiwaya}, \citenamefont {Saitoh}, \citenamefont {Kashiwaya}, \citenamefont
  {Koyanagi}, \citenamefont {Sato}, \citenamefont {Yada}, \citenamefont
  {Tanaka},\ and\ \citenamefont {Maeno}}]{Kashiwaya2019Sr2RuO4}%
  \BibitemOpen
  \bibfield  {author} {\bibinfo {author} {\bibfnamefont {S.}~\bibnamefont
  {Kashiwaya}}, \bibinfo {author} {\bibfnamefont {K.}~\bibnamefont {Saitoh}},
  \bibinfo {author} {\bibfnamefont {H.}~\bibnamefont {Kashiwaya}}, \bibinfo
  {author} {\bibfnamefont {M.}~\bibnamefont {Koyanagi}}, \bibinfo {author}
  {\bibfnamefont {M.}~\bibnamefont {Sato}}, \bibinfo {author} {\bibfnamefont
  {K.}~\bibnamefont {Yada}}, \bibinfo {author} {\bibfnamefont {Y.}~\bibnamefont
  {Tanaka}},\ and\ \bibinfo {author} {\bibfnamefont {Y.}~\bibnamefont
  {Maeno}},\ }\href {https://doi.org/10.1103/PhysRevB.100.094530} {\bibfield
  {journal} {\bibinfo  {journal} {Phys. Rev. B}\ }\textbf {\bibinfo {volume}
  {100}},\ \bibinfo {pages} {094530} (\bibinfo {year} {2019})}\BibitemShut
  {NoStop}%
\bibitem [{\citenamefont {Grinenko}\ \emph {et~al.}(2021)\citenamefont
  {Grinenko}, \citenamefont {Ghosh}, \citenamefont {Sarkar}, \citenamefont
  {Orain}, \citenamefont {Nikitin}, \citenamefont {Elender}, \citenamefont
  {Das}, \citenamefont {Guguchia}, \citenamefont {Br{\"{u}}ckner},
  \citenamefont {Barber}, \citenamefont {Park}, \citenamefont {Kikugawa},
  \citenamefont {Sokolov}, \citenamefont {Bobowski}, \citenamefont {Miyoshi},
  \citenamefont {Maeno}, \citenamefont {Mackenzie}, \citenamefont {Luetkens},
  \citenamefont {Hicks},\ and\ \citenamefont {Klauss}}]{Grinenko2019Sr2RuO4}%
  \BibitemOpen
  \bibfield  {author} {\bibinfo {author} {\bibfnamefont {V.}~\bibnamefont
  {Grinenko}}, \bibinfo {author} {\bibfnamefont {S.}~\bibnamefont {Ghosh}},
  \bibinfo {author} {\bibfnamefont {R.}~\bibnamefont {Sarkar}}, \bibinfo
  {author} {\bibfnamefont {J.-C.}\ \bibnamefont {Orain}}, \bibinfo {author}
  {\bibfnamefont {A.}~\bibnamefont {Nikitin}}, \bibinfo {author} {\bibfnamefont
  {M.}~\bibnamefont {Elender}}, \bibinfo {author} {\bibfnamefont
  {D.}~\bibnamefont {Das}}, \bibinfo {author} {\bibfnamefont {Z.}~\bibnamefont
  {Guguchia}}, \bibinfo {author} {\bibfnamefont {F.}~\bibnamefont
  {Br{\"{u}}ckner}}, \bibinfo {author} {\bibfnamefont {M.~E.}\ \bibnamefont
  {Barber}}, \bibinfo {author} {\bibfnamefont {J.}~\bibnamefont {Park}},
  \bibinfo {author} {\bibfnamefont {N.}~\bibnamefont {Kikugawa}}, \bibinfo
  {author} {\bibfnamefont {D.~A.}\ \bibnamefont {Sokolov}}, \bibinfo {author}
  {\bibfnamefont {J.~S.}\ \bibnamefont {Bobowski}}, \bibinfo {author}
  {\bibfnamefont {T.}~\bibnamefont {Miyoshi}}, \bibinfo {author} {\bibfnamefont
  {Y.}~\bibnamefont {Maeno}}, \bibinfo {author} {\bibfnamefont {A.~P.}\
  \bibnamefont {Mackenzie}}, \bibinfo {author} {\bibfnamefont {H.}~\bibnamefont
  {Luetkens}}, \bibinfo {author} {\bibfnamefont {C.~W.}\ \bibnamefont
  {Hicks}},\ and\ \bibinfo {author} {\bibfnamefont {H.-H.}\ \bibnamefont
  {Klauss}},\ }\href {https://doi.org/10.1038/s41567-021-01182-7} {\bibfield
  {journal} {\bibinfo  {journal} {Nat. Phys.}\ }\textbf {\bibinfo {volume}
  {17}},\ \bibinfo {pages} {748} (\bibinfo {year} {2021})}\BibitemShut
  {NoStop}%
\bibitem [{\citenamefont {Wilson}\ and\ \citenamefont
  {Das}(2013)}]{wilson2013}%
  \BibitemOpen
  \bibfield  {author} {\bibinfo {author} {\bibfnamefont {B.~J.}\ \bibnamefont
  {Wilson}}\ and\ \bibinfo {author} {\bibfnamefont {M.~P.}\ \bibnamefont
  {Das}},\ }\href {https://doi.org/10.1088/0953-8984/25/42/425702} {\bibfield
  {journal} {\bibinfo  {journal} {J. Phys.: Condens. Matter}\ }\textbf
  {\bibinfo {volume} {25}},\ \bibinfo {pages} {425702} (\bibinfo {year}
  {2013})}\BibitemShut {NoStop}%
\bibitem [{\citenamefont {Ghosh}\ \emph {et~al.}(2021)\citenamefont {Ghosh},
  \citenamefont {Annett},\ and\ \citenamefont {Quintanilla}}]{ghosh2021}%
  \BibitemOpen
  \bibfield  {author} {\bibinfo {author} {\bibfnamefont {S.~K.}\ \bibnamefont
  {Ghosh}}, \bibinfo {author} {\bibfnamefont {J.~F.}\ \bibnamefont {Annett}},\
  and\ \bibinfo {author} {\bibfnamefont {J.}~\bibnamefont {Quintanilla}},\
  }\href {https://doi.org/10.1088/1367-2630/ac17ba} {\bibfield  {journal}
  {\bibinfo  {journal} {New J. Phys.}\ }\textbf {\bibinfo {volume} {23}},\
  \bibinfo {pages} {083018} (\bibinfo {year} {2021})}\BibitemShut {NoStop}%
\bibitem [{\citenamefont {Ghosh}\ \emph {et~al.}(2020)\citenamefont {Ghosh},
  \citenamefont {Smidman}, \citenamefont {Shang}, \citenamefont {Annett},
  \citenamefont {Hillier}, \citenamefont {Quintanilla},\ and\ \citenamefont
  {Yuan}}]{ghosh2020}%
  \BibitemOpen
  \bibfield  {author} {\bibinfo {author} {\bibfnamefont {S.~K.}\ \bibnamefont
  {Ghosh}}, \bibinfo {author} {\bibfnamefont {M.}~\bibnamefont {Smidman}},
  \bibinfo {author} {\bibfnamefont {T.}~\bibnamefont {Shang}}, \bibinfo
  {author} {\bibfnamefont {J.~F.}\ \bibnamefont {Annett}}, \bibinfo {author}
  {\bibfnamefont {A.~D.}\ \bibnamefont {Hillier}}, \bibinfo {author}
  {\bibfnamefont {J.}~\bibnamefont {Quintanilla}},\ and\ \bibinfo {author}
  {\bibfnamefont {H.}~\bibnamefont {Yuan}},\ }\href
  {https://doi.org/10.1088/1361-648x/abaa06} {\bibfield  {journal} {\bibinfo
  {journal} {J. Phys.: Condens. Matter}\ }\textbf {\bibinfo {volume} {33}},\
  \bibinfo {pages} {033001} (\bibinfo {year} {2020})}\BibitemShut {NoStop}%
\bibitem [{\citenamefont {Mackenzie}\ and\ \citenamefont
  {Maeno}(2003)}]{mackenzie2003}%
  \BibitemOpen
  \bibfield  {author} {\bibinfo {author} {\bibfnamefont {A.~P.}\ \bibnamefont
  {Mackenzie}}\ and\ \bibinfo {author} {\bibfnamefont {Y.}~\bibnamefont
  {Maeno}},\ }\href {https://doi.org/10.1103/RevModPhys.75.657} {\bibfield
  {journal} {\bibinfo  {journal} {Rev. Mod. Phys.}\ }\textbf {\bibinfo {volume}
  {75}},\ \bibinfo {pages} {657} (\bibinfo {year} {2003})}\BibitemShut
  {NoStop}%
\bibitem [{\citenamefont {Young}\ \emph {et~al.}(2012)\citenamefont {Young},
  \citenamefont {Zaheer}, \citenamefont {Teo}, \citenamefont {Kane},
  \citenamefont {Mele},\ and\ \citenamefont {Rappe}}]{young2012}%
  \BibitemOpen
  \bibfield  {author} {\bibinfo {author} {\bibfnamefont {S.~M.}\ \bibnamefont
  {Young}}, \bibinfo {author} {\bibfnamefont {S.}~\bibnamefont {Zaheer}},
  \bibinfo {author} {\bibfnamefont {J.~C.~Y.}\ \bibnamefont {Teo}}, \bibinfo
  {author} {\bibfnamefont {C.~L.}\ \bibnamefont {Kane}}, \bibinfo {author}
  {\bibfnamefont {E.~J.}\ \bibnamefont {Mele}},\ and\ \bibinfo {author}
  {\bibfnamefont {A.~M.}\ \bibnamefont {Rappe}},\ }\href
  {https://doi.org/10.1103/PhysRevLett.108.140405} {\bibfield  {journal}
  {\bibinfo  {journal} {Phys. Rev. Lett.}\ }\textbf {\bibinfo {volume} {108}},\
  \bibinfo {pages} {140405} (\bibinfo {year} {2012})}\BibitemShut {NoStop}%
\bibitem [{\citenamefont {Yang}\ and\ \citenamefont
  {Nagaosa}(2014)}]{yang2014}%
  \BibitemOpen
  \bibfield  {author} {\bibinfo {author} {\bibfnamefont {B.-J.}\ \bibnamefont
  {Yang}}\ and\ \bibinfo {author} {\bibfnamefont {N.}~\bibnamefont {Nagaosa}},\
  }\href {https://doi.org/10.1038/ncomms5898} {\bibfield  {journal} {\bibinfo
  {journal} {Nat. Commun.}\ }\textbf {\bibinfo {volume} {5}},\ \bibinfo {pages}
  {4898} (\bibinfo {year} {2014})}\BibitemShut {NoStop}%
\bibitem [{\citenamefont {Yang}\ \emph {et~al.}(2015)\citenamefont {Yang},
  \citenamefont {Morimoto},\ and\ \citenamefont {Furusaki}}]{yang2015}%
  \BibitemOpen
  \bibfield  {author} {\bibinfo {author} {\bibfnamefont {B.-J.}\ \bibnamefont
  {Yang}}, \bibinfo {author} {\bibfnamefont {T.}~\bibnamefont {Morimoto}},\
  and\ \bibinfo {author} {\bibfnamefont {A.}~\bibnamefont {Furusaki}},\ }\href
  {https://doi.org/10.1103/PhysRevB.92.165120} {\bibfield  {journal} {\bibinfo
  {journal} {Phys. Rev. B}\ }\textbf {\bibinfo {volume} {92}},\ \bibinfo
  {pages} {165120} (\bibinfo {year} {2015})}\BibitemShut {NoStop}%
\bibitem [{\citenamefont {Ferreira}\ \emph {et~al.}(2021)\citenamefont
  {Ferreira}, \citenamefont {Manesco}, \citenamefont {Dorini}, \citenamefont
  {Correa}, \citenamefont {Weber}, \citenamefont {Machado},\ and\ \citenamefont
  {Eleno}}]{ferreira2021}%
  \BibitemOpen
  \bibfield  {author} {\bibinfo {author} {\bibfnamefont {P.~P.}\ \bibnamefont
  {Ferreira}}, \bibinfo {author} {\bibfnamefont {A.~L.~R.}\ \bibnamefont
  {Manesco}}, \bibinfo {author} {\bibfnamefont {T.~T.}\ \bibnamefont {Dorini}},
  \bibinfo {author} {\bibfnamefont {L.~E.}\ \bibnamefont {Correa}}, \bibinfo
  {author} {\bibfnamefont {G.}~\bibnamefont {Weber}}, \bibinfo {author}
  {\bibfnamefont {A.~J.~S.}\ \bibnamefont {Machado}},\ and\ \bibinfo {author}
  {\bibfnamefont {L.~T.~F.}\ \bibnamefont {Eleno}},\ }\href
  {https://doi.org/10.1103/PhysRevB.103.125134} {\bibfield  {journal} {\bibinfo
   {journal} {Phys. Rev. B}\ }\textbf {\bibinfo {volume} {103}},\ \bibinfo
  {pages} {125134} (\bibinfo {year} {2021})}\BibitemShut {NoStop}%
\bibitem [{\citenamefont {Huddart}\ \emph {et~al.}(2021)\citenamefont
  {Huddart}, \citenamefont {Onuorah}, \citenamefont {Isah}, \citenamefont
  {Bonfa}, \citenamefont {Blundell}, \citenamefont {Clark}, \citenamefont
  {De~Renzi},\ and\ \citenamefont {Lancaster}}]{huddart2021}%
  \BibitemOpen
  \bibfield  {author} {\bibinfo {author} {\bibfnamefont {B.}~\bibnamefont
  {Huddart}}, \bibinfo {author} {\bibfnamefont {I.}~\bibnamefont {Onuorah}},
  \bibinfo {author} {\bibfnamefont {M.}~\bibnamefont {Isah}}, \bibinfo {author}
  {\bibfnamefont {P.}~\bibnamefont {Bonfa}}, \bibinfo {author} {\bibfnamefont
  {S.}~\bibnamefont {Blundell}}, \bibinfo {author} {\bibfnamefont
  {S.}~\bibnamefont {Clark}}, \bibinfo {author} {\bibfnamefont
  {R.}~\bibnamefont {De~Renzi}},\ and\ \bibinfo {author} {\bibfnamefont
  {T.}~\bibnamefont {Lancaster}},\ }\href@noop {} {\bibfield  {journal}
  {\bibinfo  {journal} {Physical review letters}\ }\textbf {\bibinfo {volume}
  {127}},\ \bibinfo {pages} {237002} (\bibinfo {year} {2021})}\BibitemShut
  {NoStop}%
\bibitem [{\citenamefont {Agterberg}\ \emph {et~al.}(1999)\citenamefont
  {Agterberg}, \citenamefont {Barzykin},\ and\ \citenamefont
  {Gor’kov}}]{agterberg1999}%
  \BibitemOpen
  \bibfield  {author} {\bibinfo {author} {\bibfnamefont {D.~F.}\ \bibnamefont
  {Agterberg}}, \bibinfo {author} {\bibfnamefont {V.}~\bibnamefont
  {Barzykin}},\ and\ \bibinfo {author} {\bibfnamefont {L.~P.}\ \bibnamefont
  {Gor’kov}},\ }\href {https://doi.org/10.1103/PhysRevB.60.14868} {\bibfield
  {journal} {\bibinfo  {journal} {Phys. Rev. B}\ }\textbf {\bibinfo {volume}
  {60}},\ \bibinfo {pages} {14868} (\bibinfo {year} {1999})}\BibitemShut
  {NoStop}%
\bibitem [{\citenamefont {Mao}\ \emph {et~al.}(2003)\citenamefont {Mao},
  \citenamefont {Rosario}, \citenamefont {Nelson}, \citenamefont {Wu},
  \citenamefont {Deac}, \citenamefont {Schiffer}, \citenamefont {Liu},
  \citenamefont {He}, \citenamefont {Regan},\ and\ \citenamefont
  {Cava}}]{mao2003}%
  \BibitemOpen
  \bibfield  {author} {\bibinfo {author} {\bibfnamefont {Z.}~\bibnamefont
  {Mao}}, \bibinfo {author} {\bibfnamefont {M.}~\bibnamefont {Rosario}},
  \bibinfo {author} {\bibfnamefont {K.}~\bibnamefont {Nelson}}, \bibinfo
  {author} {\bibfnamefont {K.}~\bibnamefont {Wu}}, \bibinfo {author}
  {\bibfnamefont {I.}~\bibnamefont {Deac}}, \bibinfo {author} {\bibfnamefont
  {P.}~\bibnamefont {Schiffer}}, \bibinfo {author} {\bibfnamefont
  {Y.}~\bibnamefont {Liu}}, \bibinfo {author} {\bibfnamefont {T.}~\bibnamefont
  {He}}, \bibinfo {author} {\bibfnamefont {K.}~\bibnamefont {Regan}},\ and\
  \bibinfo {author} {\bibfnamefont {R.~J.}\ \bibnamefont {Cava}},\ }\href
  {https://doi.org/10.1103/PhysRevB.67.094502} {\bibfield  {journal} {\bibinfo
  {journal} {Phys. Rev. B}\ }\textbf {\bibinfo {volume} {67}},\ \bibinfo
  {pages} {094502} (\bibinfo {year} {2003})}\BibitemShut {NoStop}%
\bibitem [{\citenamefont {Kreisel}\ \emph {et~al.}(2013)\citenamefont
  {Kreisel}, \citenamefont {Wang}, \citenamefont {Maier}, \citenamefont
  {Hirschfeld},\ and\ \citenamefont {Scalapino}}]{kreisel2013}%
  \BibitemOpen
  \bibfield  {author} {\bibinfo {author} {\bibfnamefont {A.}~\bibnamefont
  {Kreisel}}, \bibinfo {author} {\bibfnamefont {Y.}~\bibnamefont {Wang}},
  \bibinfo {author} {\bibfnamefont {T.~A.}\ \bibnamefont {Maier}}, \bibinfo
  {author} {\bibfnamefont {P.~J.}\ \bibnamefont {Hirschfeld}},\ and\ \bibinfo
  {author} {\bibfnamefont {D.~J.}\ \bibnamefont {Scalapino}},\ }\href
  {https://doi.org/10.1103/PhysRevB.88.094522} {\bibfield  {journal} {\bibinfo
  {journal} {Phys. Rev. B}\ }\textbf {\bibinfo {volume} {88}},\ \bibinfo
  {pages} {094522} (\bibinfo {year} {2013})}\BibitemShut {NoStop}%
\bibitem [{\citenamefont {Singh}\ \emph {et~al.}(2017)\citenamefont {Singh},
  \citenamefont {Barker}, \citenamefont {Thamizhavel}, \citenamefont {Paul},
  \citenamefont {Hillier},\ and\ \citenamefont {Singh}}]{singh2017}%
  \BibitemOpen
  \bibfield  {author} {\bibinfo {author} {\bibfnamefont {D.}~\bibnamefont
  {Singh}}, \bibinfo {author} {\bibfnamefont {J.~A.~T.}\ \bibnamefont
  {Barker}}, \bibinfo {author} {\bibfnamefont {A.}~\bibnamefont {Thamizhavel}},
  \bibinfo {author} {\bibfnamefont {D.~M.}\ \bibnamefont {Paul}}, \bibinfo
  {author} {\bibfnamefont {A.~D.}\ \bibnamefont {Hillier}},\ and\ \bibinfo
  {author} {\bibfnamefont {R.~P.}\ \bibnamefont {Singh}},\ }\href
  {https://doi.org/10.1103/PhysRevB.96.180501} {\bibfield  {journal} {\bibinfo
  {journal} {Phys. Rev. B}\ }\textbf {\bibinfo {volume} {96}},\ \bibinfo
  {pages} {180501} (\bibinfo {year} {2017})}\BibitemShut {NoStop}%
\bibitem [{\citenamefont {Huang}\ \emph {et~al.}(2018)\citenamefont {Huang},
  \citenamefont {Le}, \citenamefont {Che}, \citenamefont {Yin}, \citenamefont
  {Li}, \citenamefont {Yang}, \citenamefont {Fang},\ and\ \citenamefont
  {Lu}}]{huang2018}%
  \BibitemOpen
  \bibfield  {author} {\bibinfo {author} {\bibfnamefont {Q.}~\bibnamefont
  {Huang}}, \bibinfo {author} {\bibfnamefont {T.}~\bibnamefont {Le}}, \bibinfo
  {author} {\bibfnamefont {L.}~\bibnamefont {Che}}, \bibinfo {author}
  {\bibfnamefont {L.}~\bibnamefont {Yin}}, \bibinfo {author} {\bibfnamefont
  {J.}~\bibnamefont {Li}}, \bibinfo {author} {\bibfnamefont {J.}~\bibnamefont
  {Yang}}, \bibinfo {author} {\bibfnamefont {M.}~\bibnamefont {Fang}},\ and\
  \bibinfo {author} {\bibfnamefont {X.}~\bibnamefont {Lu}},\ }\href
  {https://doi.org/10.1088/2053-1591/aae2eb} {\bibfield  {journal} {\bibinfo
  {journal} {Mater. Res. Express}\ }\textbf {\bibinfo {volume} {6}},\ \bibinfo
  {pages} {016001} (\bibinfo {year} {2018})}\BibitemShut {NoStop}%
\bibitem [{\citenamefont {Singh}\ \emph {et~al.}(2018)\citenamefont {Singh},
  \citenamefont {K.~P.}, \citenamefont {Barker}, \citenamefont {Paul},
  \citenamefont {Hillier},\ and\ \citenamefont {Singh}}]{singh2018}%
  \BibitemOpen
  \bibfield  {author} {\bibinfo {author} {\bibfnamefont {D.}~\bibnamefont
  {Singh}}, \bibinfo {author} {\bibfnamefont {S.}~\bibnamefont {K.~P.}},
  \bibinfo {author} {\bibfnamefont {J.~A.~T.}\ \bibnamefont {Barker}}, \bibinfo
  {author} {\bibfnamefont {D.~M.}\ \bibnamefont {Paul}}, \bibinfo {author}
  {\bibfnamefont {A.~D.}\ \bibnamefont {Hillier}},\ and\ \bibinfo {author}
  {\bibfnamefont {R.~P.}\ \bibnamefont {Singh}},\ }\href
  {https://doi.org/10.1103/PhysRevB.97.100505} {\bibfield  {journal} {\bibinfo
  {journal} {Phys. Rev. B}\ }\textbf {\bibinfo {volume} {97}},\ \bibinfo
  {pages} {100505} (\bibinfo {year} {2018})}\BibitemShut {NoStop}%
\bibitem [{\citenamefont {Shang}\ \emph {et~al.}(2018)\citenamefont {Shang},
  \citenamefont {Pang}, \citenamefont {Baines}, \citenamefont {Jiang},
  \citenamefont {Xie}, \citenamefont {Wang}, \citenamefont {Medarde},
  \citenamefont {Pomjakushina}, \citenamefont {Shi}, \citenamefont {Mesot},
  \citenamefont {Yuan},\ and\ \citenamefont {Shiroka}}]{shang2018}%
  \BibitemOpen
  \bibfield  {author} {\bibinfo {author} {\bibfnamefont {T.}~\bibnamefont
  {Shang}}, \bibinfo {author} {\bibfnamefont {G.~M.}\ \bibnamefont {Pang}},
  \bibinfo {author} {\bibfnamefont {C.}~\bibnamefont {Baines}}, \bibinfo
  {author} {\bibfnamefont {W.~B.}\ \bibnamefont {Jiang}}, \bibinfo {author}
  {\bibfnamefont {W.}~\bibnamefont {Xie}}, \bibinfo {author} {\bibfnamefont
  {A.}~\bibnamefont {Wang}}, \bibinfo {author} {\bibfnamefont {M.}~\bibnamefont
  {Medarde}}, \bibinfo {author} {\bibfnamefont {E.}~\bibnamefont
  {Pomjakushina}}, \bibinfo {author} {\bibfnamefont {M.}~\bibnamefont {Shi}},
  \bibinfo {author} {\bibfnamefont {J.}~\bibnamefont {Mesot}}, \bibinfo
  {author} {\bibfnamefont {H.~Q.}\ \bibnamefont {Yuan}},\ and\ \bibinfo
  {author} {\bibfnamefont {T.}~\bibnamefont {Shiroka}},\ }\href
  {https://doi.org/10.1103/PhysRevB.97.020502} {\bibfield  {journal} {\bibinfo
  {journal} {Phys. Rev. B}\ }\textbf {\bibinfo {volume} {97}},\ \bibinfo
  {pages} {020502} (\bibinfo {year} {2018})}\BibitemShut {NoStop}%
\end{thebibliography}%


%
\end{document}